\documentclass[submit]{smj}

\usepackage{color}
\usepackage{rotating}

\usepackage{subcaption}
\usepackage{hyperref}

\newcommand{\by}{\mathbf{y}}

\newcommand{\bbeta}{\boldsymbol{\beta}}

\newcommand{\boldeta}{\boldsymbol{\eta}}
\newcommand{\bmu}{\boldsymbol{\mu}}

\newcommand{\bxi}{\boldsymbol{\xi}}

\newcommand{\cN}{\mathcal{N}}
\newcommand{\cG}{\mathcal{G}}

\newcommand{\Xv}{{\mathbf{X}}}
\newcommand{\pkg}[1]{{\fontseries{b}\selectfont #1}}

\Author{Gertraud Malsiner-Walli\Affil{2}, 
		Daniela Pauger\Affil{1}, 
        and Helga Wagner\Affil{1}
}
\AuthorRunning{Gertraud Malsiner-Walli \textrm{et al.}}

\Affiliations{

\item Department of Applied Statistics, 
      Johannes Kepler University Linz,
      Austria

\item Institute for Statistics and Mathematics,
	  Vienna University for Economics and Business,
      Austria

}   

\CorrAddress{Gertraud Malsiner-Walli, 
             Institute for Statistics and Mathematics,
 WU Vienna University for Economics and Business,
Welthandelsplatz 1, 
 AT--1020 Wien, 
 Austria}

\CorrEmail{gmalsine@wu.ac.at}
\CorrPhone{(+43)\;1\;31336\;5595}
\CorrFax{--}

\Title{Effect fusion using model-based clustering}
\TitleRunning{Effect fusion using model-based clustering }

\Abstract{
In social and economic studies many of the collected variables are measured on a nominal scale, often with a large number of categories. The definition of categories is usually not unambiguous and different classification schemes using either a finer or a coarser grid are possible. Categorisation has an impact when such a variable is included as covariate in a regression model: a too fine grid will result in imprecise estimates of the corresponding effects, whereas with a too coarse grid important effects will be missed, resulting in biased effect estimates and poor predictive performance.

To achieve automatic grouping of levels with essentially the same effect, we adopt a Bayesian approach and specify the prior on the level effects as a location mixture of spiky normal components. Fusion of level effects is induced by a prior on the mixture weights which encourages empty components. Model-based clustering of the effects during MCMC sampling allows to simultaneously detect categories which have essentially the same effect size and identify variables with no effect at all. The properties of this approach are investigated in simulation studies. Finally, the method is applied to analyse effects of high-dimensional categorical predictors on income in Austria.

}


\begin{document}

\maketitle

\section{Introduction}

Researchers in medicine, social and economic sciences routinely collect  data measured on a nominal  scale  as potential predictors in regression models. The usual approach to include such categorical predictors in regression type models is to define one category as the baseline or reference category and use dummy variables for the effects of all other categories with respect to this baseline. Thus, the effect of \textit{one} categorical covariate  with  $c+1$ categories is captured by a set of $c$  regression coefficients. This leads to several issues. Including such predictors even with a
  moderate  number of categories can easily lead to a high-dimensional vector of regression coefficients. Further,  only the  subset of  observations with  a specific covariate level provides information on its effect which may result in high standard errors and unstable estimates for the   effects  of  infrequent levels. 
These issues become even more pronounced if the researcher uses a fine classification grid when categorising the data.    As often the definition of categories is  not completely dictated by subject-specific matters,  the scientist could categorise observations either finer or coarser when collecting the data.  With both strategies she/he could run into problems when  categorical variables are used as covariates in a regression model: fine categories can result in only a few subjects per category and imprecise estimates of the corresponding effects, whereas estimated effects using too coarse categories might be biased due to confounding effects of finer categories.

In order to avoid the risk of overlooking substantial differences in level effects   it would be appealing to have a method which allows to
start with a large regression model including categories on a very fine classification  grid and to obtain a sparser representation of this model during estimation.  Sparsity can be  achieved whenever the effects of  a categorical  predictor can be represented by less than $c$ regression effects.  Basically there are three different situations, where  sparsity is an issue: First,   if \textit{all} level effects are zero, the whole covariate can be excluded from the model. Second, if \textit{some} of the level effects are zero, the corresponding levels can be excluded from the model and finally if some levels have essentially the \textit{same} effect on the response, sparsity  is achieved by  fusing the effects of these levels.

Usually, sparsity in regression type models  is  achieved by applying  variable selection methods  which allow  to identify regressors with non-zero effects, i.e.~lasso \citep{tib:reg} or the elastic net \citep{zou-has:reg} in the frequentist framework
and shrinkage priors \citep{par-cas:bay, gri-bro:inf} or spike and slab priors \citep{mit-bea:bay, geo-mcc:app, ish-etal:bay} in the Bayesian framework. However, these methods are not appropriate for categorical covariates as  only  single level effects  are selected or excluded from the model. Approaches that address exclusion of a whole group of regression effects have been proposed by  \cite{chi:bay, yua-lin:mod,ram-etal:bay, kyu-etal:pen}, and recently by \cite{sim-etal:spa} but none of  these approaches  allows also for effect fusion.

For metric predictors, effect fusion can be performed  by the fused lasso \citep{tib-etal:spa} and the Bayesian fused lasso \citep{kyu-etal:pen}. Both methods assume some ordering of effects and shrink only effect differences of consecutive levels to zero and hence   are not appropriate for nominal predictors where any pair of level effects  should be subject to fusion.
Explicit  effect fusion for nominal predictors is considered in \cite{bon-rei:sim} and by Gertheiss and Tutz \citep{ger-tut:pen, ger-etal:sel, ger-tut:spa, tut-ger:reg} who specify lasso-type penalties on effects and effect differences. In a Bayesian approach, recently  \cite{pau-wag:bay} specified  a prior distribution that can be interpreted as a spike and slab prior on effects and effect differences.
However, these approaches are limited to  covariates with  a moderately large number of categories as  for a covariate with  $c+1$ categories $\binom{c+1}{2}$ possible differences have to be considered which inflates the large model even more. 


An appealing approach for effect fusion which  avoids  classification of effect differences and allows to fuse effects directly is to use model-based clustering techniques	 which  rely on mixture prior distributions.  Sparse modelling of regression effects by specifying a mixture prior is so far primarily used for continuous variables. \cite{yen-etal:var} and \cite{yen-etal:var2016} define a normal mixture prior for the regression effects and determine  the number of components, i.e.~coefficient groups, using model choice criteria.
In a nonparametric framework, \cite{mac-dun:bay} use an infinite mixture of heavy-tailed double-exponential distributions on the coefficients of continuous predictors  to allow groups of coefficients to be shrunk towards the same, possibly non-zero, mean.   
Only  \cite{dun-etal:bay} consider categorical covariates. They  propose  a multi-level Dirichlet process prior (DP) on  the effects of single nucleotide polymorphism (SNP) in genetic association studies. This prior takes the hierarchical structure of the predictors into account and allows clustering of SNPs both within and across genes. However, by considering 22 markers, each with  three levels, only a small number of levels is investigated.

Following this line of research we propose to achieve model based clustering of level effects by specifying a finite normal mixture prior. Our approach is  explicitly  designed to address effect fusion for categorical covariates and has several advantageous  features.

	 First, fusing the level effects directly   instead of   focusing on all effect differences enables us to handle categorical covariates with a large number of categories, e.g.~100 or more. 
 Second, the specified mixture prior can be interpreted as a generalisation of the standard spike and slab prior \citep{geo-mcc:var} where  a spike distribution at zero is combined with a rather flat slab distribution to allow selective shrinkage of effects, see \cite{mal-wag:com} for an overview. We replace the slab distribution by a location mixture distribution with different, non-zero means. This mixture prior allows to shrink non-zero effects to various non-zero values and introduces a natural clustering of the categories: if level effects are assigned to the same mixture  component, they are assumed to be (almost) identical and can be fused. 

 Third, the hyperparameters of the mixture prior are chosen very carefully  to achieve the modelling aims. Their specification is based on  the data to yield recommendations that
 are applicable to a wide range of real data situations. 
 The 'fineness' of the estimated level classification can be guided by the size of the specified component variance, with smaller variances inducing a larger  number of estimated effect groups. 
The  prior  on the mixture weights  is specified following  the concept of 'sparse finite mixture' \citep{mal-etal:mod}. Specifying   a sparsity inducing prior on the weights in an overfitting mixture avoids  unnecessary splitting of superfluous components  and  encourages  concentration  of  the posterior distribution on a sparse cluster solution and  thus allows to estimate the number of  effect groups from the data. 

 Fourth, remaining in the framework of finite mixture of normals and conditionally
	conjugate priors avoids a computationally intensive  estimation as standard Markov chain Monte Carlo (MCMC) methods  can be used. The MCMC scheme  for posterior inference basically combines  a regression  and a model-based clustering step,  where in  both   only standard Gibbs sampling steps are needed. 
	


 Finally, model selection consists in the  identification of the level groups and is based on the posterior draws of the partitions. Two strategies are pursued to select the final partition of the levels, by either selecting the most frequent sampled model or determining  the optimal partition of the effects based on their joint posterior fusion probabilities.



The paper is organised as follows.  In Section~\ref{sec:mod_prior}  the model and the prior distributions for the model parameters    are introduced. Details on posterior inference and model selection are given in Section~\ref{sec:inference}. The method is evaluated in a simulation study in Section~\ref{sec:sim} and applied to a regression model for   income data in  Austria in Section~\ref{sec:application}. Finally,   Section \ref{sec:discussion} concludes.

\section{Effect clustering prior} \label{sec:mod_prior}

We consider a standard linear regression model with observations $i$, $i = 1,...,N$, continuous response $y$   and $J$ categorical covariates with  categories $0, ..., c_j$ where $j=1,...,J$. For each covariate, $0$ is defined as the baseline category and $X_{jk}$ denotes the dummy variable corresponding to the $k$-th category of covariate $j$. Hence, the regression model is given as 
\begin{align}
y_i= \beta_0 + \sum_{j=1}^J \sum_{k=1}^{c_j}  X_{jk} \beta_{jk}+ \epsilon,
\label{regmod}
\end{align}
where $\epsilon \sim \mathcal{N}(0,\sigma^2)$ is a Normal  error term, $\beta_0$ is the intercept, and  $\beta_{jk}$, $k=1,\dots, c_j$ is the effect of the $k$-th category of covariate $j$ with respect to the baseline category. We call $\beta_{jk}$ the 'level effect' of category $k$.

To complete Bayesian model specification  prior distributions have to be assigned to all model parameters.  We assume that regression effects are independent between covariates and  use a prior of the structure
\begin{align}
p(\bbeta,\sigma^2) = p(\beta_0)\prod_{j=1}^J p(\bbeta_j|\xi_j) p(\sigma^2),
\label{eq:prior}
\end{align}
where $\bxi_j$ denotes additional covariate-specific hyperparameters. A flat normal prior $p(\beta_0) \sim \cN(0,B_0)$ is assigned to the intercept, and an improper inverse gamma distribution $p(\sigma^2)\sim \cG^{-1}(s_0,S_0)$ with $s_0=S_0=0$ to the error variance.\\

Our goal is to specify  a prior for  the level effects of covariate $j$   which  allows the identification of   effect groups.
Therefore, we specify  a finite mixture of normal distributions as a  prior
on the level effects $\beta_{jk}$. In contrast to the popular spike and slab priors  employed for selection of regression effects, we use a location mixture of more than two components which have a small variance, i.e.~all components are spiky. 

The prior on a regression effect $\beta_{jk}$ is specified hierarchically as
\begin{align}
p(\beta_{jk}) & =  \sum \limits_{l=0}^{L_j} \eta_{jl} f_{\mathcal{N}}(\beta_{jk}|\mu_{jl},\psi_{j}) \label{eq:mix}\\
\boldsymbol{\eta}_j & \sim   Dir_{L_j+1}(e_0) \\
\mu_{j0} & = 0 \\
\mu_{jl} & \sim \mathcal{N}(m_{j0},M_{j0})  \quad \text{for} \quad l=1,...,L_j,\label{eq:mujl}
\end{align}
where $L_j+1$ is the number of normal mixture components for covariate $j$ with location parameters $\mu_{jl}$  and  scale parameter $\psi_j$. 
For each covariate, the location parameter  of the first component $\mu_{j0}$   is fixed at 0 to allow  identification of categories which have the same  effect as the baseline category. If all level effects are assigned to this component, the covariate can be  completely  excluded from the model.  We subsume in $\bmu_j=(\mu_{j1},\dots, \mu_{jL_j})$  all other component means, 
which are assumed to be conditionally independent and follow a  flat Normal hyperprior with location and scale parameters $m_{j0}$ and $M_{j0}$. For each covariate, the variance $\psi_j$ is  the same for all components in order to ensure that each level effect group has  the same dispersion, however $\psi_j$ may  vary between covariates. Finally, a symmetric Dirichlet distribution $Dir_{L_j+1}(e_0)$  with parameter $e_0$ is specified for the mixture weights  $\boldeta_j=(\eta_{j0},\ldots,\eta_{jL_j})$.

An alternative  to our   finite mixture approach would be to specify  an  infinite mixture based on a Dirichlet process prior  $DP(\alpha)$  for  the level effects of covariate $j$. 	In this case,  the a-priori specification of the number of components  $L_j+1$  --  a well-known limitation of finite mixtures  -- would not be  necessary as it  can be estimated from the data. However, we overcome this weakness of finite mixtures by specifying a sparse finite  mixture \citep{mal-etal:mod} as prior on the level effects. This   allows  to estimate the number of 'true' components through the number of 'non-empty' components in an overfitting mixture. More details on this strategy will be provided in Section \ref{sec:hyperparameter}.
		
Additionally, it has to be pointed out that the clustering behaviour of finite and infinite mixtures is quite different. For infinite mixtures the  a priori expected number of level groups is proportional to $\alpha \cdot log(c_j)$ \citep{mac-dun:bay,mal-etal:mod} which means that with increasing number of levels $c_j$ also the number of expected clusters increases. In contrast, for a finite mixture prior as proposed here, the a-priori number of non-empty level groups is asymptotically independent of the number of levels $c_j$ \citep{mal-etal:mod}.  Hence using a finite mixture prior for the effects of a categorical predictor seems more suitable, as  one would expect that  in a hierarchical categorisation scheme there exists a certain level of aggregation which is able to capture all relevant effect differences and   the number of different effect sizes would not increase with a 'finer' classification grid. 
		
		%

\subsection{Choice of hyperparameters}\label{sec:hyperparameter}

The specification of the prior hyperparameters is crucial to achieve our modelling aims. To obtain  recommendations  that  are applicable to a wide range of situations, we take an empirical approach and choose the hyperparameters depending on  the data.  
 
The location parameter of the first mixture component $\mu_{j0}$ is fixed at  $0$ in order to allow fusion to the baseline. For the location parameters of all other components $\mu_{jk}$, we specify  a 
normal hyperprior  located at the 'centre' of the effects and with  large variance  in order to induce  only little shrinkage to the prior mean. Thus, we set  the mean $m_{0j}$ of the normal hyperprior to $m_{0j}=mean({\hat{\bbeta_j}})$  and the variance $M_{0j}$ to the squared range of $\hat{\bbeta}_j$, i.e.~$M_{0j}=(\max_{k} \hat{\beta}_{jk} -\min _{k} \hat{\beta}_{jk})^2$, where $\hat{\bbeta}_j$ is the estimated coefficient vector of covariate $j$ under flat prior.
 
Levels effects should be assigned to the same component only if the  sizes  of their effects are almost identical. Therefore, specification of the component variance $\psi_j$ is crucial as it  reflects the notion of \textit{negligible/relevant} effect differences. As the prior on the component variance $\psi_j$ should take  into account the  scaling of covariates, we allow $\psi_j$ to vary across covariates but not between levels of  one covariate. 
 
We  define the component variance $\psi_j$ as some proportion $1/\nu$  from 
 the variation of the estimated level effects $\hat{\bbeta}_j$ under  flat  prior, i.e.~$\psi_j=\frac{1}{\nu}V_j$, where $V_j =\frac{1}{c_j-1} \sum_{k=1}^{c_j}(\hat{\beta}_{jk}-\bar{\beta}_j)^2 $ and $\bar{\beta}_j=\frac{1}{c_j} \sum_{k=1}^{c_j} \hat{\beta}_{jk}$.
 With increasing $\nu$  the shapes of the mixture components become more spiky and more distinct  groups of level effects will be identified. Thus, $\psi_j$ implicitly  controls  the 'fineness' of the estimated  partition of level effects and  hence the size of the selected model. As mentioned above,  the component variances are defined covariate-specific in order to account for the dispersion of the level estimates within a covariate. However, the component variances could  also be  specified globally, i.e.~with the same spike size for all covariates,  if interest lies in defining a 'global' threshold for level effect differences across all covariates. 
 
 %
 
\begin{figure}[!ht]\centering
 	\centering
 	\includegraphics[width=0.48\columnwidth]{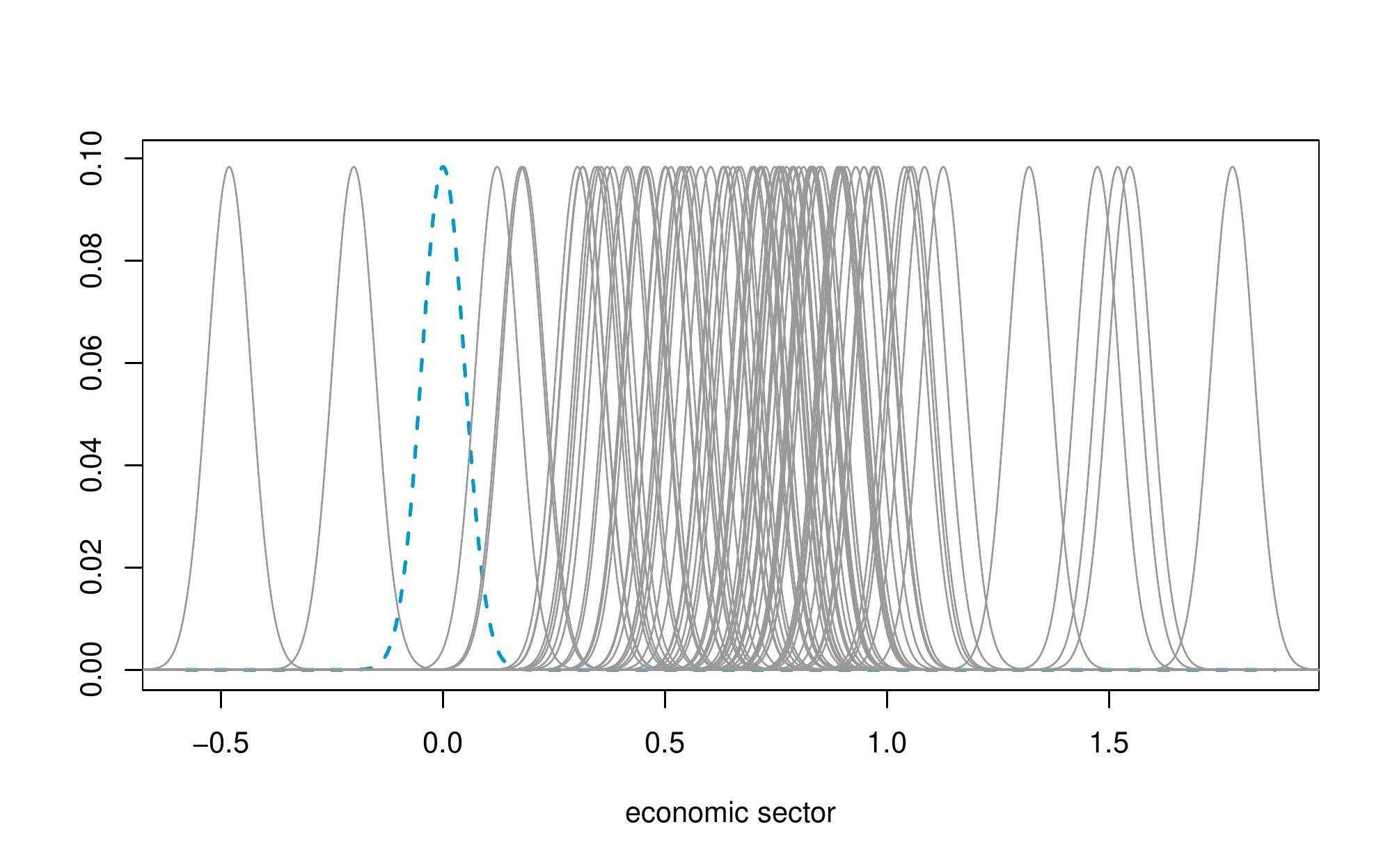}
 	 \hfill
 	\includegraphics[width=0.48\columnwidth]{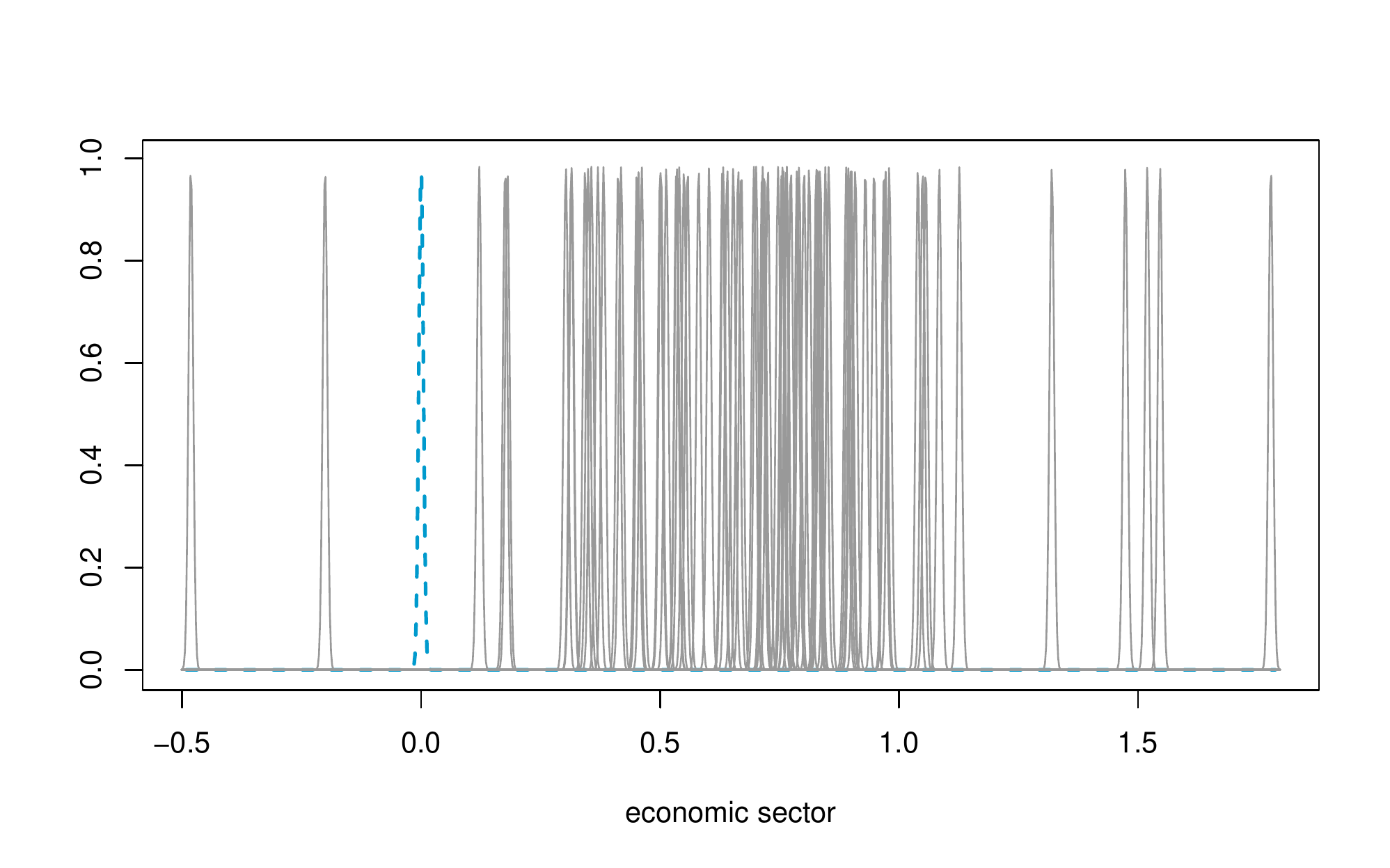} 
 	\caption{\label{fig:prior} Finite mixture prior on level effects of covariate 
 		{\tt economic sector} for two different mixture component variances, $\nu=10^2$ (left panel) and $\nu=10^4$ (right panel). One component is centred at zero (blue dashed line), the others at $\hat{\beta}_{jk}$, $k=1,\dots c_j,$ under   flat prior.   
 	}
 \end{figure}

Figure~\ref{fig:prior} shows the prior distributions of  the level effects of one of the covariates 
 in our application, the  covariate {\tt economic sector} with 84 levels,  for two values of the component variance $\psi_j$. One mixture component is centred at zero and the others at the posterior means   $\hat{\beta}_{jk}$ under a  standard flat  Normal prior.  
 
Since  the choice of the prior  component variance  $\psi_j$  influences effect fusion, as an alternative we consider $\psi_j$ to be random  with   a hyperprior   $\psi_j \sim \mathcal{G}^{-1}(g_0, G_{0j})$. We expect to obtain  more robust cluster solutions as the influence of a fixed parameter $\psi_j$ should be mitigated. For a given value  of $g_0$, we choose $G_{0j}$ such that the a priori expected component variance  $E(\psi_j)=\frac{G_{0j}}{g_0-1}$ matches a desired size, i.e.~$E(\psi_j) \approx \frac{V_j}{\nu}$, and hence  set  $G_{0j}=\frac{V_j}{\nu}(g_0-1)$. As the variance  is  given as $V(\psi_j)=E(\psi_j)^2/(g_0-2)$, the scale parameter $g_0$ controls the deviation from the expected value. To allow only for small deviations from the expectation we set $g_0=100$. Thus, a priori the standard deviation for $\psi_j$ is around $1/10$ of the expected mean. We investigate the influence of the variance parameter $\nu$  for  fixed variance $\psi_j$ as well as under a hyperprior in the simulation study in Section \ref{sec:sim}.
 
We now turn to  the specification of the number of mixture components  $L_j+1$. We set $L_j=c_j$ in order to capture the redundant case where all effects are different from each other (and from the baseline). Thus, our prior  defines  an overfitting mixture model, where the mixture distribution on the level effects has more components than level effects to be estimated. In order to achieve a sparser estimation of the overfitting mixture model by encouraging superfluous components to be emptied we follow \cite{mal-etal:mod}, who base their approach on \cite{rou-men:asy}.

 		
\cite{rou-men:asy} investigated the asymptotic behaviour of the posterior distribution of an overfitting mixture model and showed that the hyperparameter $e_0$ of the Dirichlet prior on the mixture  weights determines whether superfluous components will be   left empty or split in two or more identical components.  Asymptotically, if $e_0 < d/2$, where $d$ is the dimension of the component-specific parameter,  the posterior expectation of the weights  converges to zero for superfluous components. In contrast,  for $e_0 > d/2$ the posterior distribution handles overfitting by defining at least two identical components with non-negligible weights. Hence, in order  to encourage  empty components in the overfitting mixture prior for the level effects, we specify  
a sparsity inducing prior on the mixture weights $\boldsymbol{\eta}_j$ with   $e_0 <d/2$, where  $d=2$ is the dimension of $(\mu_{kj},\psi_j)$. Then, superfluous mixture components should be emptied during MCMC sampling and the sampled partitions concentrate on the model space with sparse solutions.  Following  \cite{mal-etal:mod}, we choose $e_0$ very small, e.g.~$e_0 = 0.01$, to actually empty all superfluous components, also in covariates with many levels.

\section{Posterior inference} \label{sec:inference}
The posterior distribution, which results from combining the likelihood derived from equation (\ref{regmod})  with  the prior distribution of $(\bbeta,\sigma^2)$ specified in (\ref{eq:prior}) - (\ref{eq:mujl}),
is not of closed form and therefore  MCMC methods are used for posterior inference.
During  MCMC sampling the whole model space will be explored, i.e.~different clustering solutions for the covariate effects will be visited, which allows to assess model uncertainty and also
to determine model averaged estimates.

However, though model averaged estimates of the coefficients may give good results in terms of prediction, researchers are often interested  in selection of a \textit{final} model and interpretation of its results. In regression models with categorical predictors, model selection is more involved than in  standard variable selection, as the problem is  to determine an appropriate clustering of level effects, which means that both the number of clusters as well as  the members of each cluster have to be determined. We address this problem in Section \ref{sec:model_sel}, where we present two different strategies for  model selection for the effects of a categorical covariate.

\subsection{MCMC sampling}\label{sec:MCMC}

Parameter estimation is performed through MCMC sampling based on data augmentation \citep{Mix:Diebolt+Robert:1994,Mix:Fruehwirth2006}. For each covariate $j$, latent allocation variables $\boldsymbol{S}_j = S_{j1},...,S_{jc_j}$ are introduced to indicate the component a regression effect $\beta_{jk}$ is assigned to. $S_{jk}$ takes values in $\{0,1,\ldots,L_j\}$. Conditional on $S_{jk}=l$, the prior distribution for $\beta_{jk}$ is the normal mixture component distribution
$$ \beta_{jk}|S_{jk}=l \sim \mathcal{N}(\mu_{jl},\psi_{j}).$$ MCMC sampling is basically performed by iterating two steps: the regression step, where the level effects and the  error variance are sampled conditional on knowing the components the effects are assigned to, and the model-based clustering step, where the parameters of the mixture components and the latent allocation variables are sampled. In the starting configuration,
each level effect $\beta_{jk}$ is assigned to  a  separate component $l$,   where both the component mean and the  effect are estimated under  flat prior. The component located at zero is left empty. 

The MCMC sampling scheme iterates the following steps:
\begin{itemize}
	\item[] Regression steps
	\begin{enumerate}
		\item Sample  the regression coefficients $\bbeta$ conditional on the latent allocation variable $\boldsymbol{S}$ from the Normal posterior $\mathcal{N}(\boldsymbol{b}_N, \boldsymbol{B}_N)$.  
		\item Sample the error variance $\sigma^2$ from its full conditional posterior distribution $\mathcal{G}^{-1}(s_N, S_N)$.
	\end{enumerate}
	\item[] Model based clustering steps
	\begin{enumerate}
		\setcounter{enumi}{2}
		\item For $j=1,...,J$ sample the component weights $\boldeta_j$ from the Dirichlet distribution $Dir(e_{j0},e_{j1},\ldots,e_{jL_j})$.
		\item For $j=1,...,J$; $l=1,\dots L_j$ sample the mixture component means $\mu_{jl}$ from their Normal posterior $\mathcal{N}(m_{jl},M_{jl})$.
		\item If a hyperprior is  specified  on ${\psi_j}$,  sample the mixture component variances $\psi_{j}$ from their inverse gamma posterior $\mathcal{G}^{-1}(g_{jN},G_{jN})$ for $j=1,...,J$;  otherwise this step is omitted.
		\item  Sample the latent allocation indicators $\boldsymbol{S}$ from their full conditional posterior 
		$$ P(S_{jk}=l|\beta_{jk},\boldsymbol{\mu}_j,\psi_j) \propto \eta_{jl} f_{\mathcal{N}}(\beta_{jk}|\mu_{jl},\psi_{j}).$$
	\end{enumerate}
\end{itemize}
More details on the sampling steps are given in Appendix \ref{app:MCMC}. The method is implemented in the {\tt R} package {\tt effectFusion} \citep{pau-etal:eff} which is available on CRAN.

\subsection{Model averaged estimates}
MCMC  draws approximate  the  whole posterior distribution  taking  into account model uncertainty:  e.g.~for a regression effect 
$\beta_{jk}$ the posterior is the mixture  distribution
\[  p(\beta_{jk}|\by)=\sum_{i}^{} p(\beta_{jk}|\by,\mathcal{M}^{(i)})p(\mathcal{M}^{(i)}|\by).\]
where the mixture components are model-specific  posterior distributions 
and the mixture weights are  the posterior model probabilities.  
Hence, the mean over all MCMC draws for  $\beta_{jk}$ should be a robust, model-averaged  estimator. Its  predictive performance is investigated in Section \ref{sec:sim}.

\subsection{Model selection} \label{sec:model_sel}
To perform model selection,  generally  the sampled mixture models have to be identified.
In the Bayesian framework, identification of a finite mixture model requires handling the 'label switching' problem \citep{red-wal:mix} which is caused by the invariance of representation (\ref{eq:mix}) with respect to reordering the components:
\begin{align*}
p(\beta_{jk})&= \sum_{l=0}^{L_j} \eta_{jl} \;f_\mathcal{N}(\beta_{jk}|\mu_{jl},\psi_j)\\
&= \sum_{l=0}^{L_j} \eta_{j\rho(l)} \;f_\mathcal{N}(\beta_{j\rho(l)}|\mu_{j\rho(l)},\psi_j),
\end{align*} 
where $\rho$ is an arbitrary permutation of $\{0,\ldots,L_j\}$.  Practically, it may happen, that during MCMC sampling the labels associated with the components change,  which impedes component-specific inference from the MCMC output. The label switching problem is usually solved by post-processing the MCMC output in order to obtain a unique labelling of the draws. We avoid the label switching problem by basing model selection on the information  whether a  pair of level effects is assigned to the same or to different clusters. For each iteration $m$ and  each covariate $j$, we construct the $(L_j+1) \times (L_j+1)$ matrix $\boldsymbol{M}_j^{(m)}$ with entry 1  if the two corresponding levels $g$ and $h$ belong to the same cluster, and 0 otherwise, i.e.
$$M_{j,gh}^{(m)}=I_{\{S_{jg}^{(m)}=S_{jh}^{(m)}\}}.$$ 
This matrix is independent of the component labelling and therefore invariant to label switching. It  contains the  clustering information for covariate $j$, i.e.~all information regarding number of effect groups and group memberships. 

After MCMC sampling, there are several options to summarise the posterior clustering distribution and to select a final partition of the level effects of covariate $j$. One possibility is   to choose the partition $\boldsymbol{M}_j$ that was selected most often during MCMC sampling. Since the  parameter $e_0$ of the Dirichlet distribution  is specified very small, according to \cite{rou-men:asy} 'true' clusters should not be split.  The posterior distribution will  concentrate on  parsimonious  partitions of the  effects  and the number of clusters will  depend only  on the specified spike variance size.  Thus, the posterior mode estimate, i.e  the model sampled  most frequently during MCMC sampling should be a good 
choice   for the final model.    

Another option to select the final partition 
is to average the matrix $\boldsymbol{M}_j^{(m)}$ over all $N_m$ MCMC iterations yielding the matrix $\boldsymbol{C}_j=\frac{1}{N_m}\sum_{m=1}^{N_m}\boldsymbol{M}_j^{(m)}$. Its entries 
$C_{j,gh}$ correspond to the relative frequency with which effects of   two levels $g$ and $h$ are  assigned to the same cluster and approximate the  posterior probability that $\beta_{jg}$ and   $\beta_{jh}$
are members of the same cluster. Hence, each matrix  $\boldsymbol{C}_j$ can be interpreted as a 'similarity' matrix: a value of $C_{j,gh}$ close to $1$ indicates that the two level effects are almost identical.
To find a clustering of the levels effects which corresponds most closely  to the similarity matrix, 
we follow \cite{mol-etal:bay} and use $k$-medoids clustering.

Similar to $k$-means clustering, $k$-medoids clustering aims at clustering points by minimising the distances between points assigned to a cluster and the point defined as the centre of the cluster. $k$-medoids  always chooses a data point as centre of a cluster ('medoid') and works with arbitrary distance metrics  between the data points. This feature makes it attractive for our approach since the similarity matrix can easily be transformed to a distance matrix   $\mathbf{D}_j=\mathbf{1}-\mathbf{C}_j$, where $\mathbf{1}$ is a matrix with elements  $1$.
We use the clustering algorithm Partitioning Around Medoids (PAM) proposed by \cite{kau-rou:fin} which yields  an optimal partition for a specified  number of clusters. 
The final partition is chosen by comparing partitions
with different numbers of clusters by  their silhouette coefficients  \citep{rou:sil}. The definition of the silhouette coefficient is given in Appendix \ref{app:silhouette}.

An advantage of  this approach  is that level effect clusters are correctly identified even if distances are high, i.e.~joint inclusion probabilities are rather small. This can happen if the number of categories is large  and  the strong  overlapping of the mixture components induces a frequent  switching  of the levels between the components, so that the inclusion probability of any two level effects become small, and the most frequent model is not a good representative of the sampled models. However, a  drawback of this approach is that the silhouette coefficient can not be computed for a one-cluster solution. Therefore,  with this  strategy it is not possible to identify the case where all level effects are assigned to the zero  component and the corresponding predictor can be  excluded from the model.

\section{Simulation study} \label{sec:sim}

A sparser  representation of the effects of a categorical covariate is possible  when 1) \textit{some} or 2) \textit{all} of the levels have no effect  at all or 3) some levels have the same effect and hence can be fused. To investigate  the  performance of the proposed prior distribution  in these situations, we perform a simulation study   where  categorical covariates with moderate as well as large number of levels represent the various types of sparsity. 
We evaluate both  model selection strategies proposed in Section \ref{sec:model_sel}, i.e.~using either the most frequent sampled  partition or the partition selected by performing PAM and the silhouette coefficient, with respect to correct model selection. Further, we determine  estimation accuracy and  predictive performance  of  the estimates  based on the selected models as well as the model averaged estimates.

\subsection{Set-up}
 We define a regression model according to (\ref{regmod})  with four independent categorical predictors, the first three predictors having  10 and the forth  100 categories. All categories have uniform prior class probabilities. The level effects of the first covariate have three different values  ($\bbeta_1=(0,0,0,0.5,0.5,0.5,1,1,1)$), for  the second covariate only one level has a non-zero effect on the outcome ($\bbeta_2=(0,0,0,0,0,0,0,0,1)$), the levels of the third variable have no effect at all,  and levels  of  the last covariate  has  six different effects ($0,0.5,1,1.5,2,2.5$) equally distributed among the levels. The intercept $\beta_0$ is set to zero.
100 data sets, each  consisting of $n=4,000$ observations, a  random design matrix and   a  Normal error  $\varepsilon \sim \mathcal{N}(0,0.5)$, are generated. The regression model  with prior  specifications as described in Section \ref{sec:hyperparameter} and flat prior on the intercept is fitted to the data sets.  In order to investigate the influence of the component variance, the simulations are performed with varying sizes of the  variance parameter $\nu$, i.e.~$\nu=10,10^2,\ldots,10^6$, and fixed  as well as  random  component variance specifications.

MCMC sampling is run for 15,000  iterations after a burn-in of 15,000. The final model is chosen by employing both model selection strategies  suggested in Section \ref{sec:model_sel}. The selected models are then  refitted under  a flat Normal prior $\mathcal{N}(0, \boldsymbol{I} B_0)$ with $B_0 = 10000$ on all level effects. For the refit,  MCMC is run for 3,000 iterations after a burn-in of 1,000.

In order to compare the different  final models, two model choice criteria, the Deviance Information Criterion (DIC), proposed by \cite{Cat:Spiegelhalter2002}, and the BICmcmc, suggested by \cite{Mix:Fruehwirth2011}, are performed. Both measures rely on the MCMC output and can be easily computed.
BICmcmc is determined from the largest log-likelihood value observed across the MCMC draws. Whereas the classical BIC is independent from the prior, BICmcmc depends also on the prior of the regression parameters. 

\subsection{Model selection results}

The model selection results are evaluated by reporting the estimated number of level  effect groups. Additionally, the clustering quality is assessed by calculating the adjusted Rand index \citep{hub-ara:com}, the error rate, false negative and false positive rate.

The adjusted Rand index \citep{hub-ara:com} allows to quantify the similarity between the true and estimated partition of the level effects. It is a corrected form of the Rand index \citep{ran:obj}, adjusted for chance agreement. 
A value of  $1$ corresponds to perfect agreement between two partitions whereas an adjusted Rand index of $0$ corresponds to results no better than expected by randomly drawing two partitions, each with a fixed number of clusters and a fixed number of elements in each cluster. A formal definition of the index can be found in Appendix \ref{app:randindex}. 
 
The error rate (err) of the clustering result is the number of misclassified categories divided by all categories.  It should be as small as possible. Since interest mainly  lies  in avoiding incorrect fusion of categories rather than unnecessary  splitting of 'true' groups, additionally false negative rate   (FNR)  and false positive rate  (FPR) are reported. They are  defined as
 $$  FNR = \frac{FN}{TP+FN}\qquad FPR = \frac{FP}{TN+FP},$$
where $FN$ is the number of levels incorrectly fused, $FP$ is the number of levels incorrectly split, and $TN$ and $TP$ are the number of levels fused and split  correctly, respectively.

Table \ref{tab:ev_res} shows  the clustering results for all four covariates using both model selection strategies, i.e.~the most frequent model ('most') and the model selected using PAM ('pam'), for fixed component variance $\psi_j$ and $\nu=10^3$. 'Freq' reports the  number of iterations (out of 15,000) where the most frequent model is sampled, and 'groups' reports the   estimated number of clusters. All results are averaged over 100 data sets. Obviously, 
sparsity is achieved  for all covariates. The true number of clusters is correctly identified  for both strategies,  except for   covariate 3, where 'pam' is not able to select the one-cluster solution with all level effects being $0$. Also 'most' has some difficulty to fuse all levels to the baseline. However, using a broader variance by setting $\nu$ to $10$ or $10^2$, fusion to the baseline is perfect for this variable, as can be seen in Table \ref{tab:ev_res_var1} in the Appendix.
The  selected partitions under both model selection strategies show   high  values of  AR  and  low error rate indicating that the  identified clusters capture  the true group structure of level effects well. Notably,  fusion is  almost perfect also for the 100 categories of covariate $4$, with  an  average error rate of  $\text{err}=0.04$.
 
\begin{table}[h]
\centering
\small
\begin{tabular}{|c|c|ccc|cc|cc|cc|cc|}
  \hline
  Variable & freq & \multicolumn{3}{c|}{groups}  & \multicolumn{2}{c|}{AR} & \multicolumn{2}{c|}{err} & \multicolumn{2}{c|}{FPR} & \multicolumn{2}{c|}{FNR} \\ 
  &&true& most & pam & most & pam & most & pam & most & pam & most & pam \\
  \hline
   1 &14844&3 &3.0& 3.0 & 1.00 & 1.00 & 0.00 & 0.00 & 0.00 & 0.00 & 0.00 & 0.00 \\
   2&14970& 2&2.0  & 2.0 & 1.00 & 0.99 & 0.00 & 0.00 & 0.00 & 0.00 & 0.00 & 0.00 \\ 
   3&  11897 &1& 1.8 & 2.0 & 0.26 & 0.00 & 0.23 & 0.28 & 0.33 & 0.41 & - & - \\  
   4&11044 & 6&6.0 & 6.2 & 0.91 & 0.90 & 0.04 & 0.04 & 0.08 & 0.08 & 0.02 & 0.02 \\  
   \hline
\end{tabular}
\caption{ Model selection results for fixed $\psi$ with $\nu=10^3$. Comparison of the two model selection strategies 'most'  and 'pam'. The first three variables have 10 categories, the 4th variable 100 categories. FNR is not defined for variable 3. } 
\label{tab:ev_res}
\end{table}

In order to compare our clustering results to those obtained following the approach proposed by  \cite{ger-tut:spa} and \cite{oel-etal:reg},  we use the \textsf{R} package \pkg{gvcm.cat} to fit a regression model  with a regularising penalty term on the level effect differences. The penalty  parameter is chosen via  cross-validation. 
Table \ref{tab:ev_gvcm} reports the classification results.
The approach yields    large models  where   level effects  are fused very cautiously, resulting in small AR and FNR values and high values for error rate  and FPR.

\begin{table}[ht]
	\centering
	\begin{tabular}{|c|cc|cccc|}
		\hline
		Var & true & groups & AR & err & FPR & FNR  \\ 
		\hline
		1 &  3 &  9.0 & 0.12 & 0.60 & 0.91 & 0.00  \\ 
		2 &  2 &  7.7 & 0.03 & 0.66 & 0.92 & 0.00 \\ 
		3 &  1 &  7.2 & 0.00 & 0.74 & 0.92 & - \\ 
		4 &  6 & 59.8 & 0.05 & 0.83 & 0.96 & 0.01 \\ 
		\hline
	\end{tabular}
	\caption{Penalty approach: Model selection results.} 
	\label{tab:ev_gvcm}
\end{table}

To investigate the impact of  the  component variances $\psi_j$ on model selection, we ran MCMC  for various values of $\nu$ for fixed as well as random component variance $\psi_j$. In Table \ref{tab:ev_res_var4}  we report the results for covariate 4 which is of special interest  due to its large number of levels, results for all other covariates are reported in Appendix \ref{app:sim_res}. For fixed $\psi_j$, as expected, the number of identified groups increases with $\nu$  as the spike variance $\psi_j$ decreases. 
 To  detect  the `true' effect clusters, a good choice for  $\nu$ is a value in the range of  $\nu=10^2$ to  $\nu=10^3$,  also AR and error rate are good for this choice. Larger values of $\nu$ lead to a finer classification of the level effects. The number of estimated  effect groups increases up to $20$ for the very small spike variance ($\nu=10^6$), with $AR=0.46$ and error rate $0.50$. However, the relatively high values of FPR and  low values of  FNR indicate that   groups are split into subgroups while almost no levels of truly different groups are combined to  new groups.

 With  a hyperprior  on the component variances specified as described in Section \ref{sec:hyperparameter},  the true number  of effects  is captured well
for variables 1 to 3, where  the true number of clusters is at most  three,  see Tables \ref{tab:ev_res_var1}, \ref{tab:ev_res_var2}, \ref{tab:ev_res_var3} in the Appendix.  However, for covariate 4 with six different effects,   the true number of effect groups  is underestimated  using    both model selection strategies. The number of estimated components is  almost constant in  $\nu$,  particularly for large values,  
 where  more (splitted) groups would be  expected. This results suggests that a hyperprior on the component variance cannot be recommended, if a larger number of level  effect groups is expected. 

Table \ref{tab:modChoice} 
shows  that   all models  with fixed or random component variance   outperform  the full model  with respect to the BICmcmc;  models with a fixed   component variance outperform the full model even in terms of DIC  unless the component variance  is large (i.e. for $\nu=10$). 
Thus,  with  a 'reasonable' variance, i.e. $\nu$ between $10^2$ and $10^5$,  
 a good  fit of the models   can be obtained  with a  small number of coefficients.


\begin{small}
\begin{table}[ht]
	\centering
	\begin{tabular}{|c|c|c|cc|cc|cc|cc|cc|}
		\hline
		&$\nu$ & freq&  \multicolumn{2}{c|}{groups}  & \multicolumn{2}{c|}{AR} & \multicolumn{2}{c|}{Error} & \multicolumn{2}{c|}{FPR} & \multicolumn{2}{c|}{FNR} \\  
		&  &  & most & pam  & most & pam & most & pam & most & pam & most & pam\\ 
		\hline
fixed & 10 &  9 & 3.8 & 3.4 & 0.52 & 0.53 & 0.49 & 0.49 & 0.06 & 0.04 & 0.21 & 0.20 \\ 
 & $10^2$ & 54 & 6.1 & 6.1 & 0.91 & 0.93 & 0.04 & 0.03 & 0.07 & 0.06 & 0.02 & 0.01 \\ 
 & $10^3$ & 11044 & 6.0 & 6.2 & 0.91 & 0.90 & 0.04 & 0.04 & 0.08 & 0.08 & 0.02 & 0.02 \\ 
 & $10^4$ & 8077 & 6.7 & 7.0 & 0.87 & 0.86 & 0.09 & 0.09 & 0.14 & 0.15 & 0.02 & 0.01 \\ 
 & $10^5$ & 7159 & 11.0 & 11.6 & 0.69 & 0.68 & 0.28 & 0.30 & 0.40 & 0.42 & 0.01 & 0.01 \\ 
 & $10^6$ & 6800 & 19.6 & 20.6 & 0.46 & 0.44 & 0.50 & 0.53 & 0.65 & 0.68 & 0.00 & 0.00 \\ 
		\hline
random & 10 &  9 & 3.7 & 3.4 & 0.52 & 0.53 & 0.49 & 0.49 & 0.06 & 0.04 & 0.21 & 0.20 \\ 
 & $10^2$ & 46 & 4.2 & 4.3 & 0.62 & 0.64 & 0.35 & 0.30 & 0.07 & 0.10 & 0.14 & 0.12 \\ 
 & $10^3$ & 38 & 4.7 & 4.8 & 0.70 & 0.73 & 0.26 & 0.21 & 0.06 & 0.09 & 0.11 & 0.08 \\ 
 & $10^4$ & 42 & 4.9 & 4.9 & 0.73 & 0.73 & 0.24 & 0.21 & 0.06 & 0.09 & 0.09 & 0.08 \\ 
 & $10^5$ & 44 & 4.8 & 4.9 & 0.72 & 0.73 & 0.25 & 0.21 & 0.05 & 0.09 & 0.10 & 0.08 \\ 
 & $10^6$ & 45 & 4.8 & 4.9 & 0.72 & 0.74 & 0.24 & 0.20 & 0.06 & 0.09 & 0.10 & 0.08 \\
 \hline 
	\end{tabular}
	\caption{Model selection results: \textbf{Var4}, $100$ categories, true number of groups is $6$.} 
	\label{tab:ev_res_var4}
\end{table}

\begin{table}[ht]
	\centering
	\begin{tabular}{|c|c|cc|cc|}
		\hline
		& $\nu$ &\multicolumn{2}{c|}{BICmcmc} & \multicolumn{2}{c|}{DIC} \\
		&& most& pam & most& pam \\
		\hline
		true&  &\multicolumn{2}{c|}{8501} & \multicolumn{2}{c|}{8445} \\
		\hline \hline
		fixed&$10^1$  & 9154 & 9089 & 9043 & 9113\\
		&$10^2$ &  8644 & 8643 & 8579 & 8587 \\
		&$10^3$ &  8643 & 8645 & 8581 & 8582 \\
		&$10^4$ &  8646 & 8648 & 8570 & 8571 \\
		&$10^5$ &  8655 & 8657 & 8535 & 8536 \\
		&$10^6$ &  8735 & 8731 & 8527 & 8527 \\
		\hline
		random & $10$ & 9154 & 9091 & 9046 & 9114 \\
		& $10^2$ & 8915 & 8850 & 8799 & 8864 \\
		& $10^3$ & 8836 & 8771 & 8716 & 8782 \\
		& $10^4$ & 8814 & 8768 & 8713 & 8759 \\
		& $10^5$ & 8826 & 8768 & 8713 & 8772 \\
		& $10^6$ & 8826 & 8768 & 8713 & 8772 \\
		\hline
		\hline
		penalty&  &\multicolumn{2}{c|}{9365} & \multicolumn{2}{c|}{8692}\\
		\hline \hline
		full&  &\multicolumn{2}{c|}{9579} & \multicolumn{2}{c|}{8703}\\
		\hline	
	\end{tabular}
	\caption{Model choice criteria for the selected models using the model selection strategies 'most' and 'pam', the penalty approach ('penalty'), and fitting the true and the full model under flat prior.} 
	\label{tab:modChoice}
\end{table}

%

\end{small}

\clearpage
\subsection{Parameter estimation accuracy and predictive performance}
To evaluate  the performance of the proposed approach  with respect to   estimation accuracy of the parameters we compute the   mean squared error (MSE) of the coefficient estimates  by  averaging over all data set-specific mean squared errors 
$$MSE^{i}=  \frac{1}{C+1} ( (\beta_0^{true}-\hat{\beta}_0^i)^2 +  \sum \limits_{j=1}^J \sum \limits_{k=1}^{c_j} (\beta_{jk}^{true}-\hat{\beta}_{jk}^i)'(\beta_{jk}^{true}-\hat{\beta}_{jk}^i)),  \quad i=1,\dots, 100,$$  where $i$ is the number of the data set and  $C=\sum_{j=1}^{J}c_j$ is the dimension of  the vector of regression coefficients $\bbeta$ in the full model.

\begin{figure}
	\centering
	\includegraphics[scale=0.35]{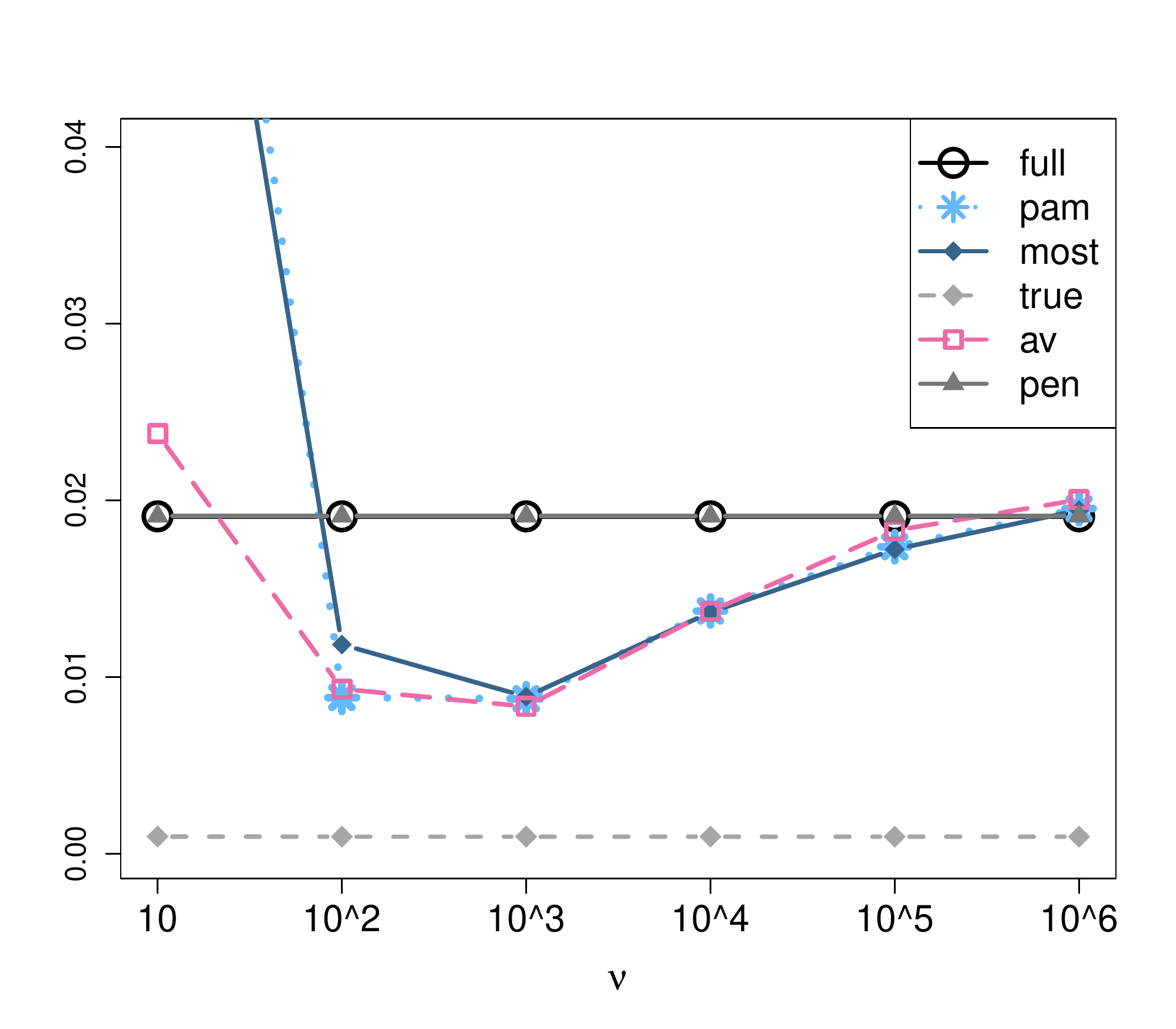}
	\includegraphics[scale=0.35]{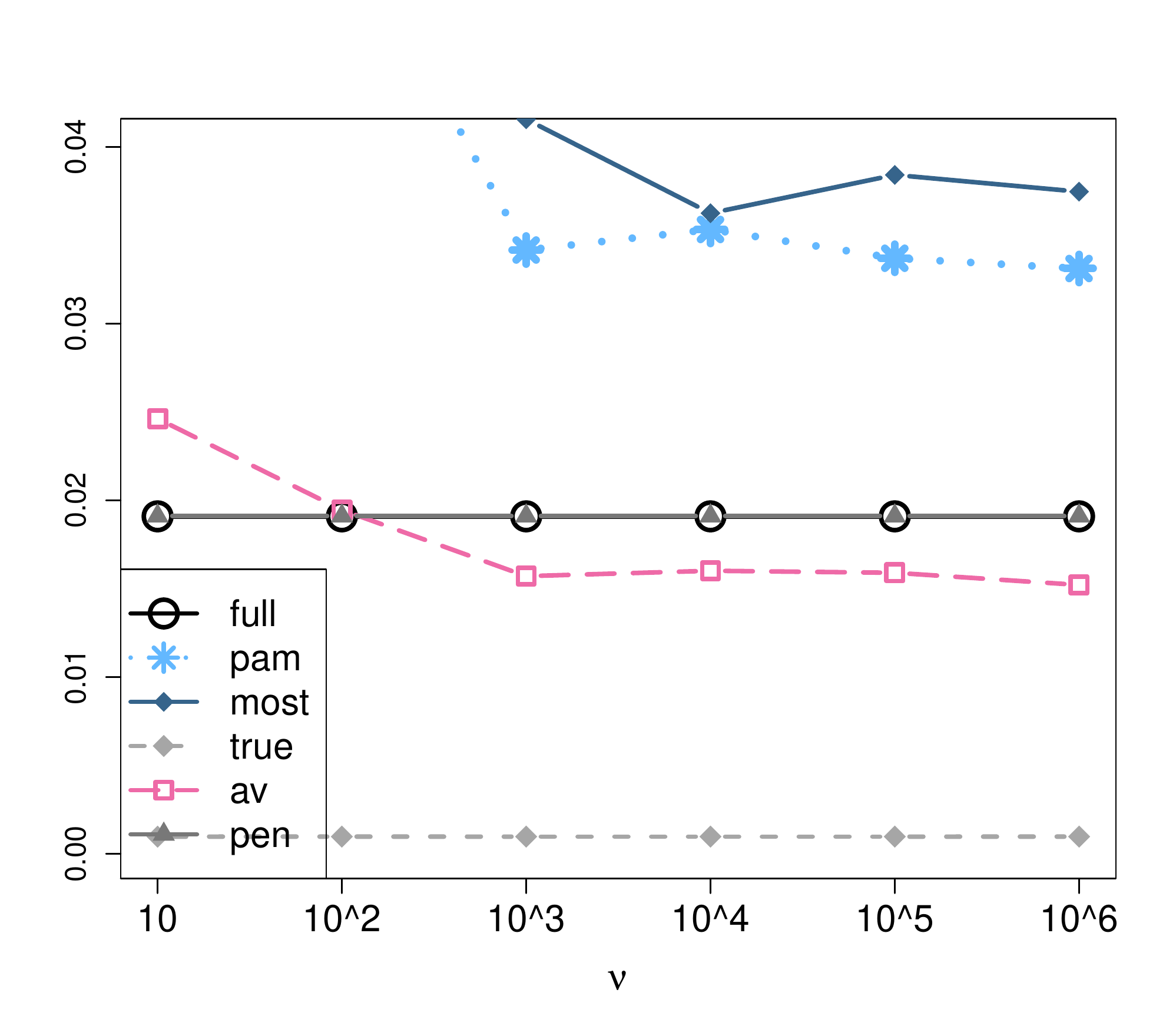} \\
    \caption{\label{fig:mse_sim} Simulation study:  Mean squared error (MSE) of coefficient estimates for various values of $\nu$ with fixed component variance $\psi_j$ (left) and hyperprior on $\psi_j$ (right), averaged over 100 simulated data sets.}
\end{figure}

In Figure  \ref{fig:mse_sim}  the  MSE  of the parameter estimates based on both model selection strategies    as well as  the  MSE for the model averaged estimates (`av') are shown for different values of $\psi_j$, and fixed and random spike variances. For comparison, also the  MSE of  the  penalized ML-estimates (`pen') and the estimates of the   full model  (`full') with a distinct effect for each level, and the true model (`true') with correctly  fused levels, both under a flat Normal prior,  are  shown.

For a fixed spike variance (plot on the left-hand side),    the MSE    for the selected models under both strategies is lower than for the full model and  penalized regression.   MSE is lowest for  $\nu=10^3$ and  increases with   larger $\nu$, but  never exceeds the  MSE of the full model.
   Notable, the model averaged coefficient estimates ('av'), which do not rely on the selection of a specific model but rather average over all sampled models, outperform the full model estimates under flat prior for each $\nu$ specification. 
   Even if the hyperprior on the the variance is specified (plot on the right-hand side) and the estimates of the selected models are worse than the full model (due to the sparse estimation of level groups in variable four, see Table \ref{tab:ev_res_var4}), the averaged estimates have smaller MSE than the full model.
    This indicates that the proposed mixture prior can also be used as an alternative to a non-informative prior in standard regression analysis, when just    accurate parameter estimation (and not model selection) is the aim of the analysis and more robust results in regard to prior specifications are desired.
         
 Finally, to investigate the predictive performance of our approach, we generate  100 new data sets $(\boldsymbol{y}^{new},\Xv^{new})$ with $N^{new}=1,000$ observations and  compute predictions of the  response vector $\boldsymbol{y}^{new}$  based on $\Xv^{new}$ and the estimates of each of the 100 original data sets.  The mean squared predictive error (MSPE)   is computed as average  of  
$$MSPE ^{i}= \frac{1}{N^{new}}  (\boldsymbol{y}^{new}-\Xv^{new}\hat{\bbeta}^i) '(\boldsymbol{y}^{new}-\Xv^{new}\hat{\bbeta}^i)$$
where $\hat{\bbeta}^i$ is the estimate in data set $i$. The average MSPE is  displayed  in Figure \ref{fig:mspe_sim} for the different estimates.  
For fixed component variances, predictions  from the selected models under the effect fusion prior (`most'  and `pam')  and also  using the model averaged estimates (`av') outperform those using the estimates from the full model and the regularised estimates (`pen'). For estimates from the selected  models with a   hyperprior on the component variance the  MSPE is larger than for those from  the full model. Note, that again model averaged estimates perform well  yielding smaller  prediction errors for values of  $\nu> 10$  thus  outperforming    estimates from  the full model.  

\begin{figure}
	\centering
	\includegraphics[scale=0.35]{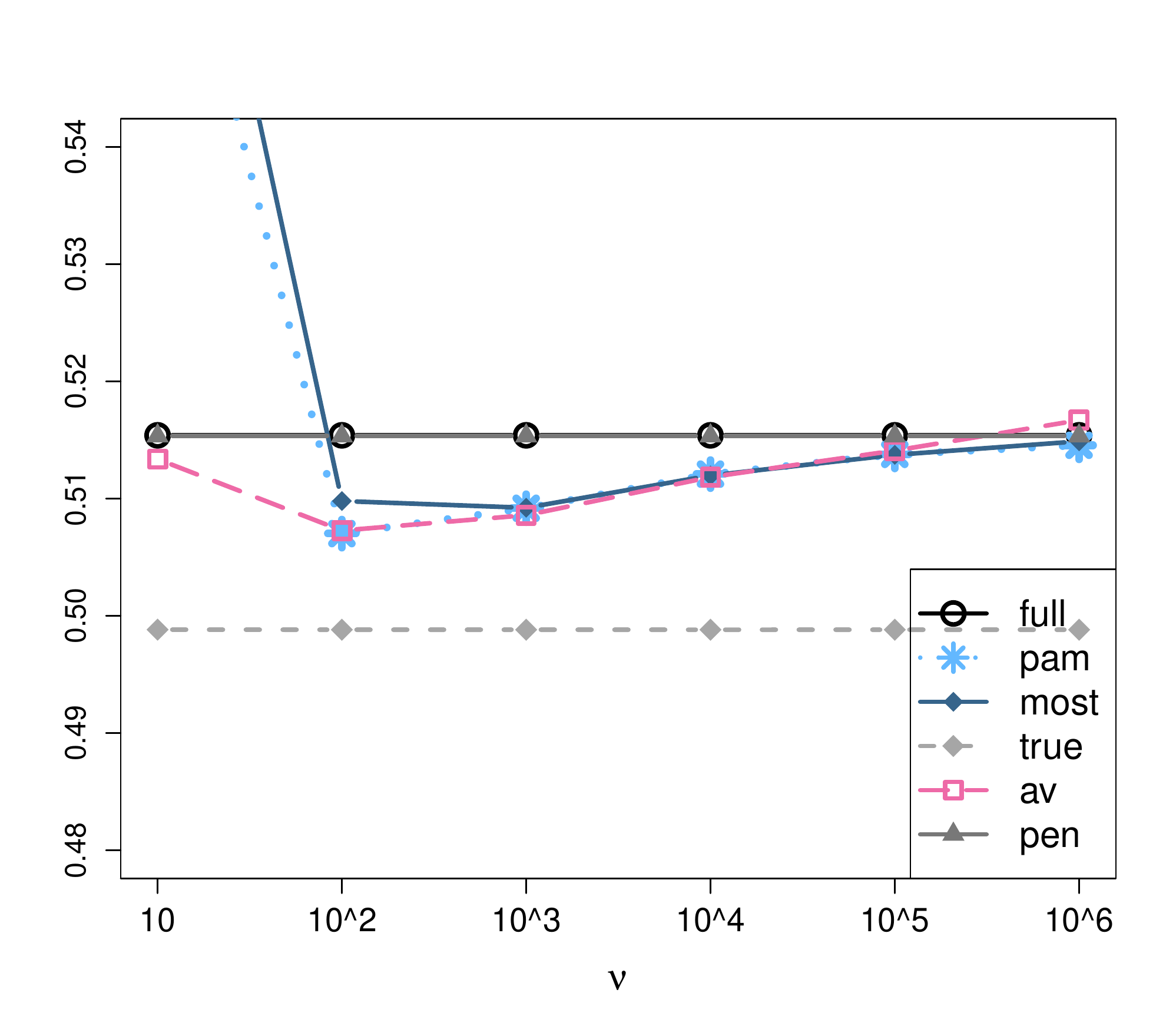}
	\includegraphics[scale=0.35]{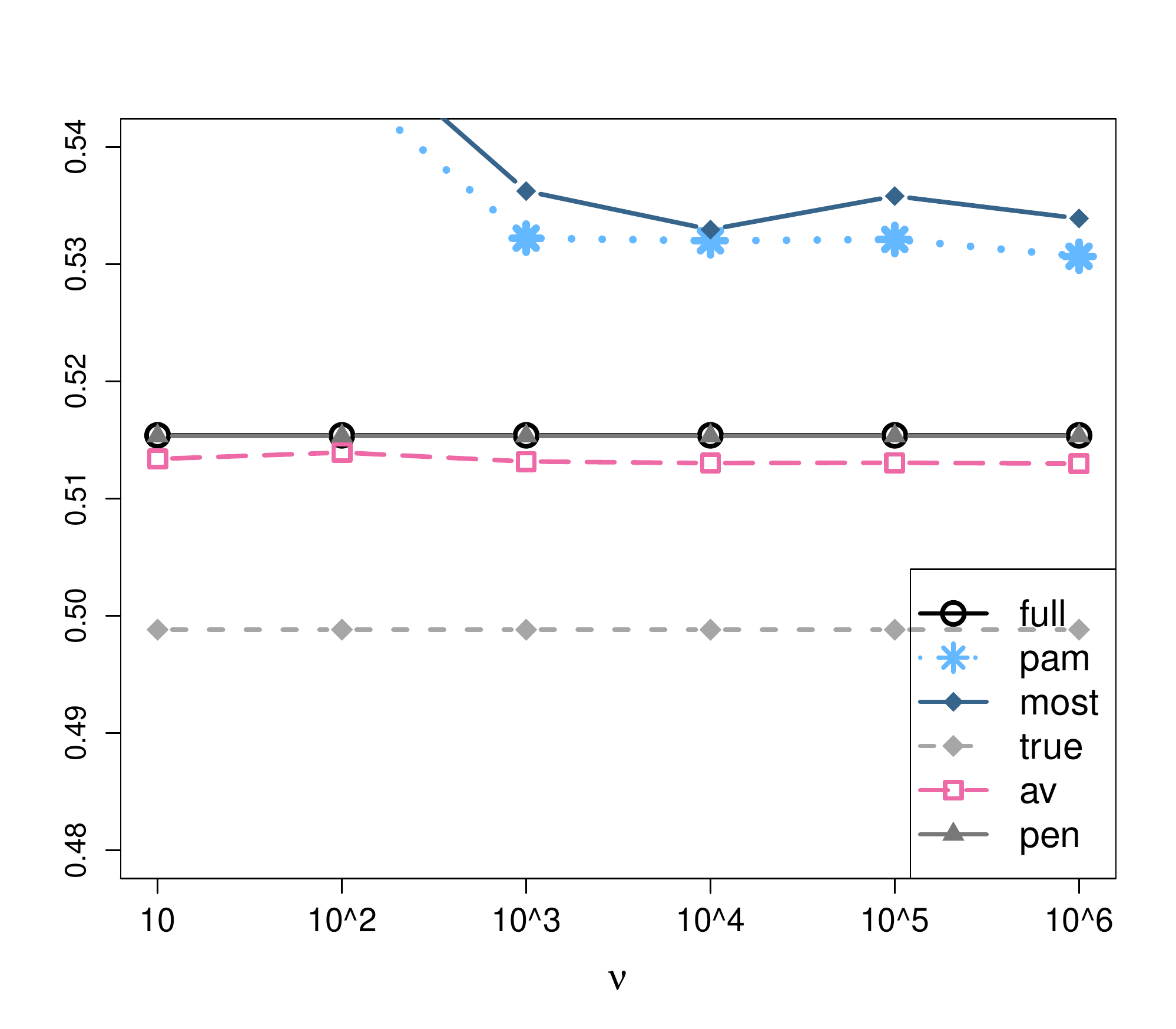} \\
    \caption{\label{fig:mspe_sim} Simulation study:  Mean squared  prediction error (MSPE) of coefficient estimates for various values of $\nu$ with fixed component variance $\psi_j$ (left) and hyperprior on $\psi_j$ (right), averaged over 100 simulated data sets.} \label{fig:mspe_sim}
\end{figure}

\section{Application} \label{sec:application}

We illustrate  the proposed  approach  for effect fusion in an application to data from  EU-SILC data set (= Statistics on Income and Living Conditions) 2010 in Austria. Relying on a questionnaire, the EU-SILC data are the main source for statistics on income distribution and social inclusion at the European level, see
 \href{http://statistik.at/web_de/frageboegen/private_haushalte/eu_silc/index.html}{web page of Statistics Austria}. We use a  linear regression model to analyse the effects of socio-demographic variables on the (log-transformed) annual income and aim at  identifying levels of  categorical covariates which account for income differences. 

As potential regressors we consider the continuous covariate {\tt age} (as linear and squared term) and the categorical predictors {\tt  gender}, {\tt citizenship}, {\tt  federal state} of residence in Austria, highest {\tt education level} a person achieved,  the economic  {\tt  sector}  a person is working  and the  {\tt job function}.
	  
The economic {\tt sector}  is classified using the classification scheme NACE (statistical classification of economic activities in the European Community), whereas {\tt job function} is determined by using a two-level scheme. Both classifications have a hierarchical structure with 21 and 5 categories on the first level and 84 and 25 categories on the  second level of aggregation, respectively. The definition of the categories of the two levels of the covariate {\tt job function} are given in more detail in Table \ref{tab:jobFunction} of  Appendix \ref{app:silc_res}.

\begin{figure}[!ht]\centering
	\includegraphics[width=0.32\textwidth]{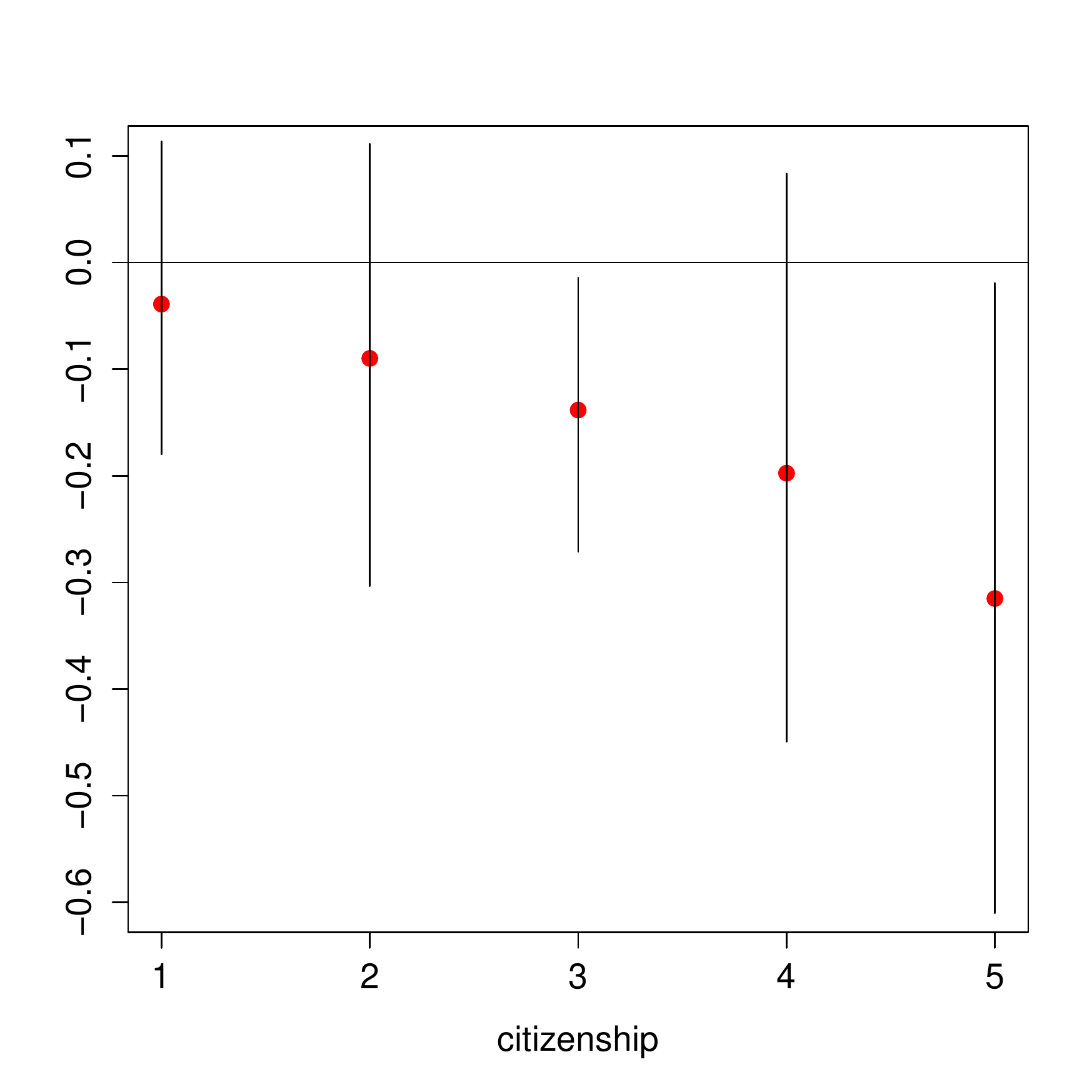} 
	\includegraphics[width=0.32\textwidth]{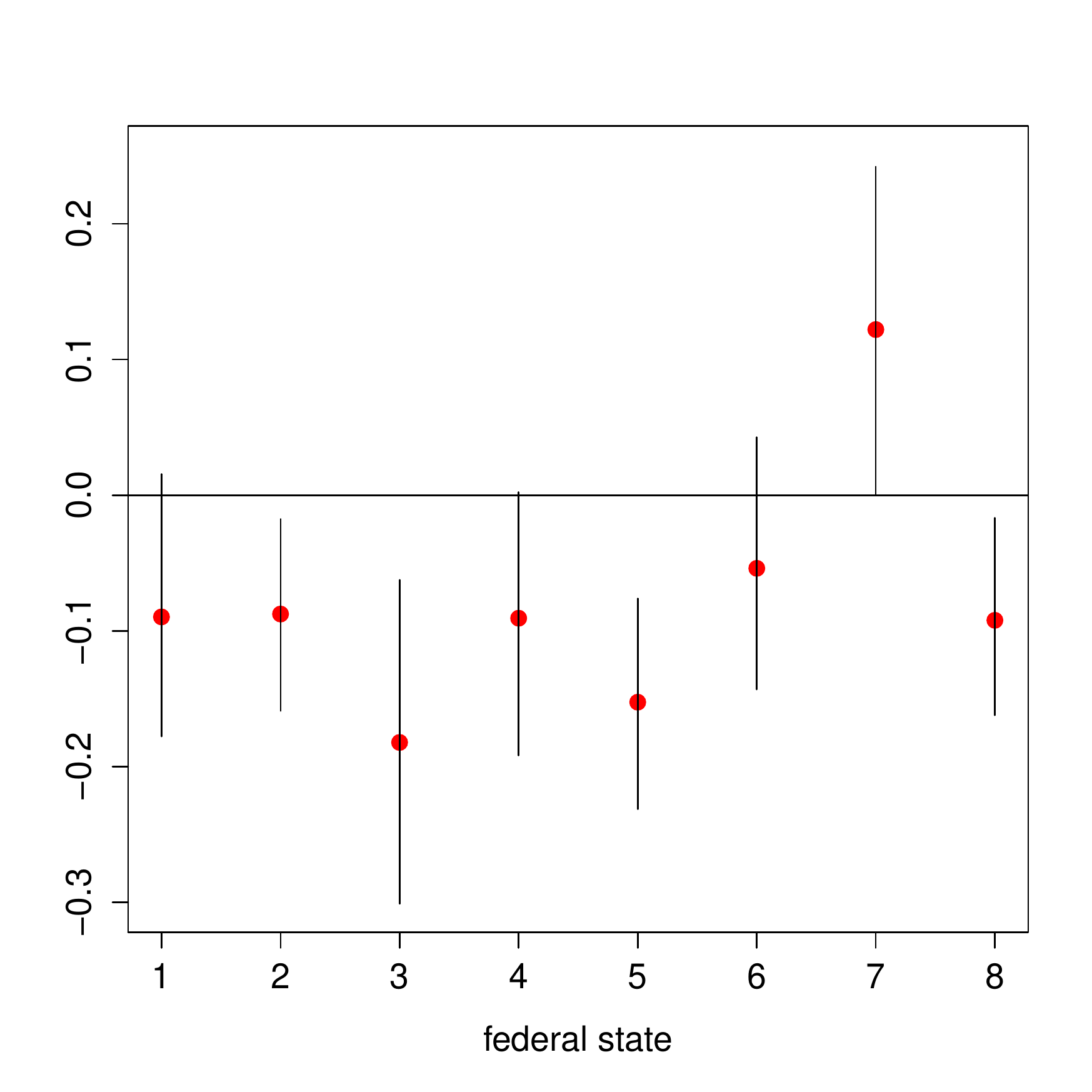} 
	\includegraphics[width=0.32\textwidth]{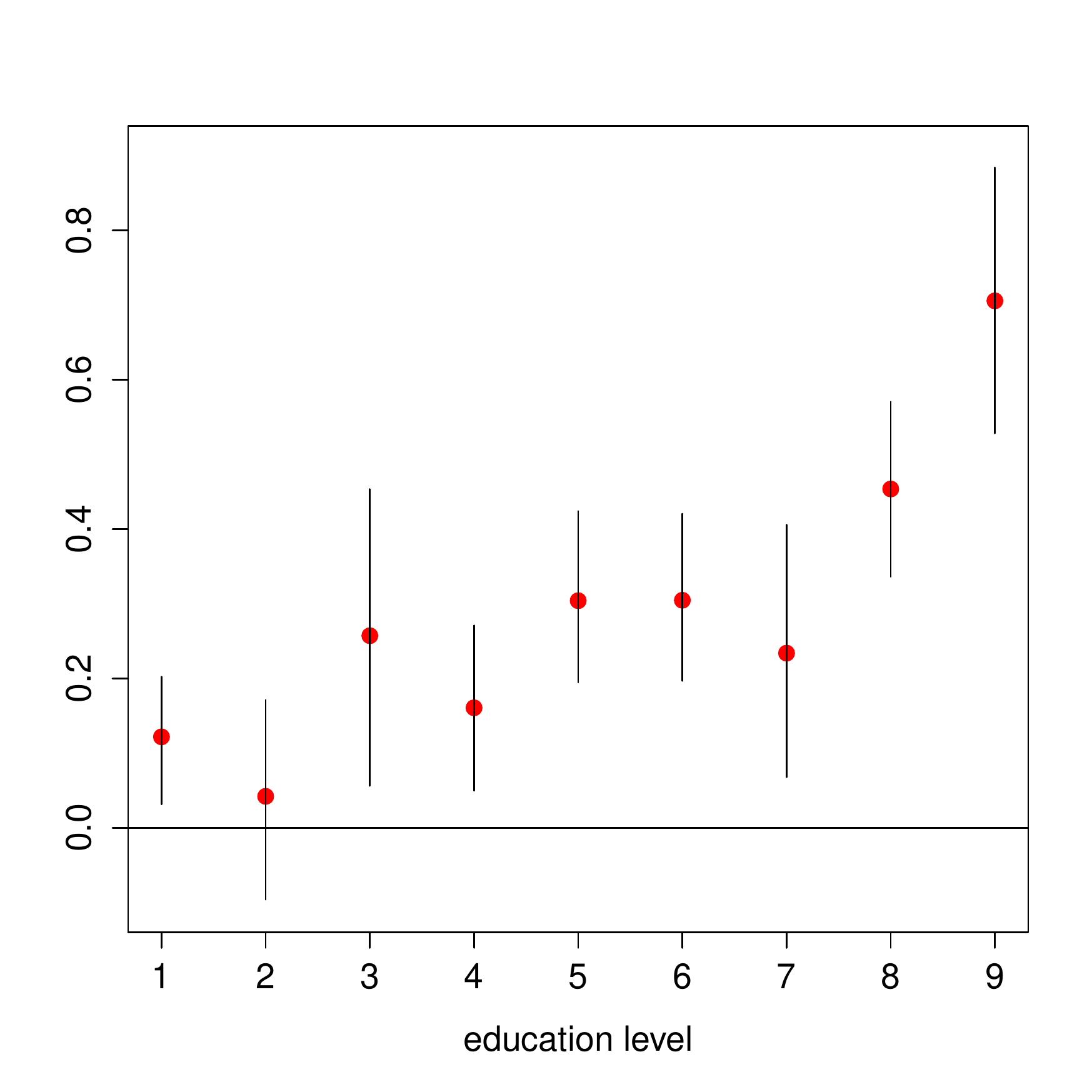} \\
	\includegraphics[width=0.32\textwidth]{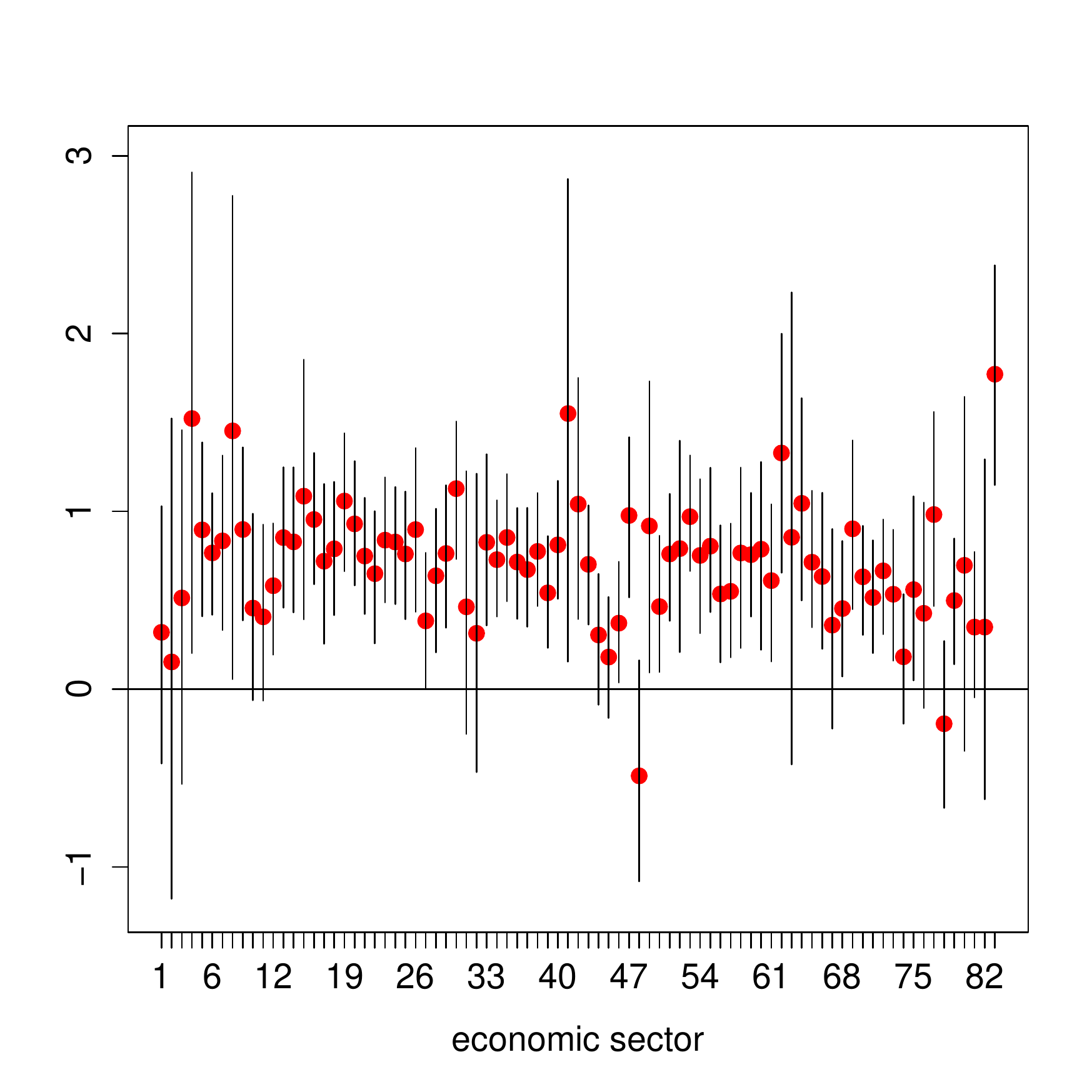} 
	\includegraphics[width=0.32\textwidth]{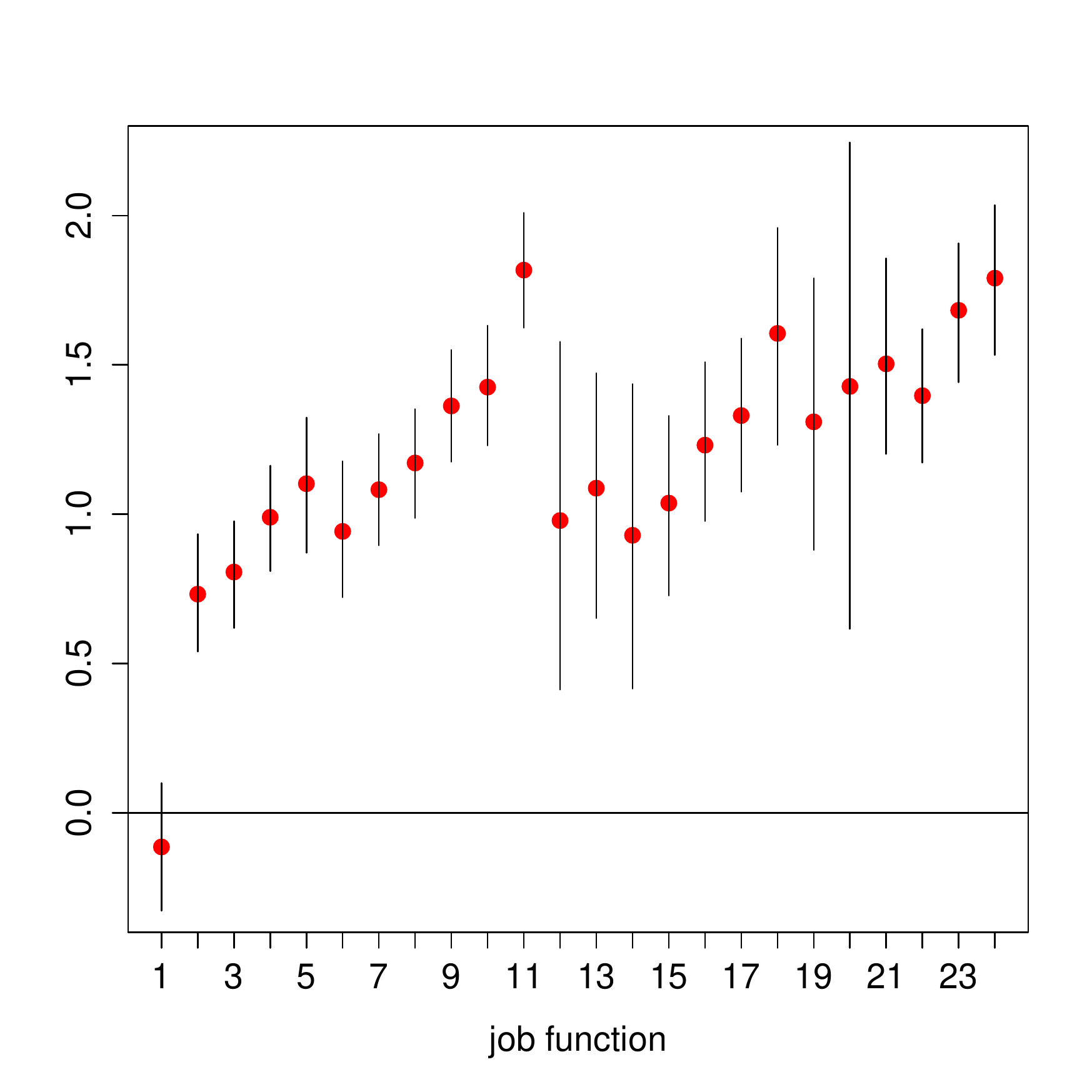} 
	\caption{SILC data, posterior means and  95\% HPD intervals under flat prior.}	
	\label{fig:silc_model_full}
\end{figure}

We  use the finer second level of aggregation and specify the effect fusion prior on the categories to achieve a sparser representation of the effects.	We standardise the response $y$ and restrict the analysis to observations of full time employees with a minimum annual income of EUR 2,000. After removing observations with missing values in the response and/or the predictors the data set consists of observations from 3,865 people. As baseline categories we choose      the categories with  the lowest labels in the classification schemes, except    {\tt
 federal state} where  the baseline is {\tt Upper Austria}. Figure \ref{fig:silc_model_full} shows  the 95\% HPD intervals for the level effects
under a flat Normal prior.

We fit  regression models with prior specifications as described in Section \ref{sec:mod_prior}, with fixed and random component variances,  $\nu=10,\ldots,10^6$, and perform model selection as described in Section \ref{sec:model_sel}. MCMC sampling is run for 15,000 iterations after a burn-in of 25,000 iterations. Table \ref{tab:silc}  reports the estimated number of level effects for each of the categorical covariates  under both model selection strategies and  for the different variance specifications. Also the results for  fitting a regularized regression using {\tt gvcm.cat} are reported. Finally, in order to evaluate the selected models, the BICmcmc of the refitted models is shown.

\begin{table}
	\centering
	\footnotesize
	\begin{tabular}{|c|c|cc|cc|cc|cc|cc|cc|}
		\hline
		&$\nu$  & \multicolumn{2}{c|}{citizen}  & \multicolumn{2}{c|}{federal state}&  \multicolumn{2}{c|}{education}&\multicolumn{2}{c|}{sector} & \multicolumn{2}{c|}{job function} &  \multicolumn{2}{c|}{$BICmcmc$} \\
		& & most & pam& most & pam& most & pam  & most & pam & most & pam & most & pam   \\
		\hline
		fixed & $10^1$ & 2 & 2 & 2 & 2 & 2 & 2 & 3 & 2 & 4 & 3 & 8774 & 8520 \\
		& $10^2$ & 2 & 2 & 2 & 2 & 3 & 3 & 3 & 2 & 5 & 5 & 8294 & 8275 \\
		& $10^3$ & 1 & 3 & 3 & 2 & 3 & 5 & 4 & 4 & 5 & 5 & 8193 & 8165 \\
		& $10^4$ & 2 & 2 & 2 & 2 & 5 & 5 & 7 & 7 & 6 & 6 & 8114 & 8117 \\
		& $10^5$ & 2 & 2 & 4 & 4 & 5 & 5 & 13 & 13 & 8 & 8 & 8174 & 8171 \\
		& $10^6$ & 3 & 3 & 4 & 4 & 6 & 6 & 17 & 20 & 11 & 11 & 8231 & 8255 \\	
		\hline
		random & $10^1$ & 2 & 2 & 2 & 2 & 2 & 2 & 3 & 2 & 3 & 2 & 9204 & 9189 \\
		& $10^2$ & 2 & 2 & 2 & 2 & 3 & 3 & 3 & 2 & 4 & 3 & 8621 & 8451 \\
		& $10^3$ & 2 & 2 & 2 & 2 & 3 & 3 & 3 & 2 & 4 & 3 & 8563 & 8451 \\
		& $10^4$ & 2 & 2 & 3 & 3 & 3 & 3 & 3 & 3 & 4 & 3 & 8559 & 8448 \\
		& $10^5$ & 2 & 2 & 3 & 3 & 3 & 3 & 3 & 2 & 4 & 3 & 8559 & 8440\\
		& $10^6$ & 2 & 2 & 3 & 3 & 3 & 3 & 3 & 2 & 4 & 3 & 8558 & 8440 \\
		\hline
		pen & -- & \multicolumn{2}{c|}{5}& \multicolumn{2}{c|}{8}& \multicolumn{2}{c|}{9}&\multicolumn{2}{c|}{7}&  \multicolumn{2}{c|}{21} &  \multicolumn{2}{c|}{8445}\\
		full & -- & \multicolumn{2}{c|}{6}& \multicolumn{2}{c|}{9}& \multicolumn{2}{c|}{10}& \multicolumn{2}{c|}{84}&  \multicolumn{2}{c|}{25} & \multicolumn{2}{c|}{9047} \\
		\hline
	\end{tabular}
	\caption{\label{tab:silc} SILC data: estimated number of level groups for the categorical covariates  and BICmcmc for various scaling factors  $\nu$, with fixed and random component variances $\psi_j$.}
\end{table}

\begin{figure}[!ht]\centering
	\includegraphics[width=0.32\textwidth]{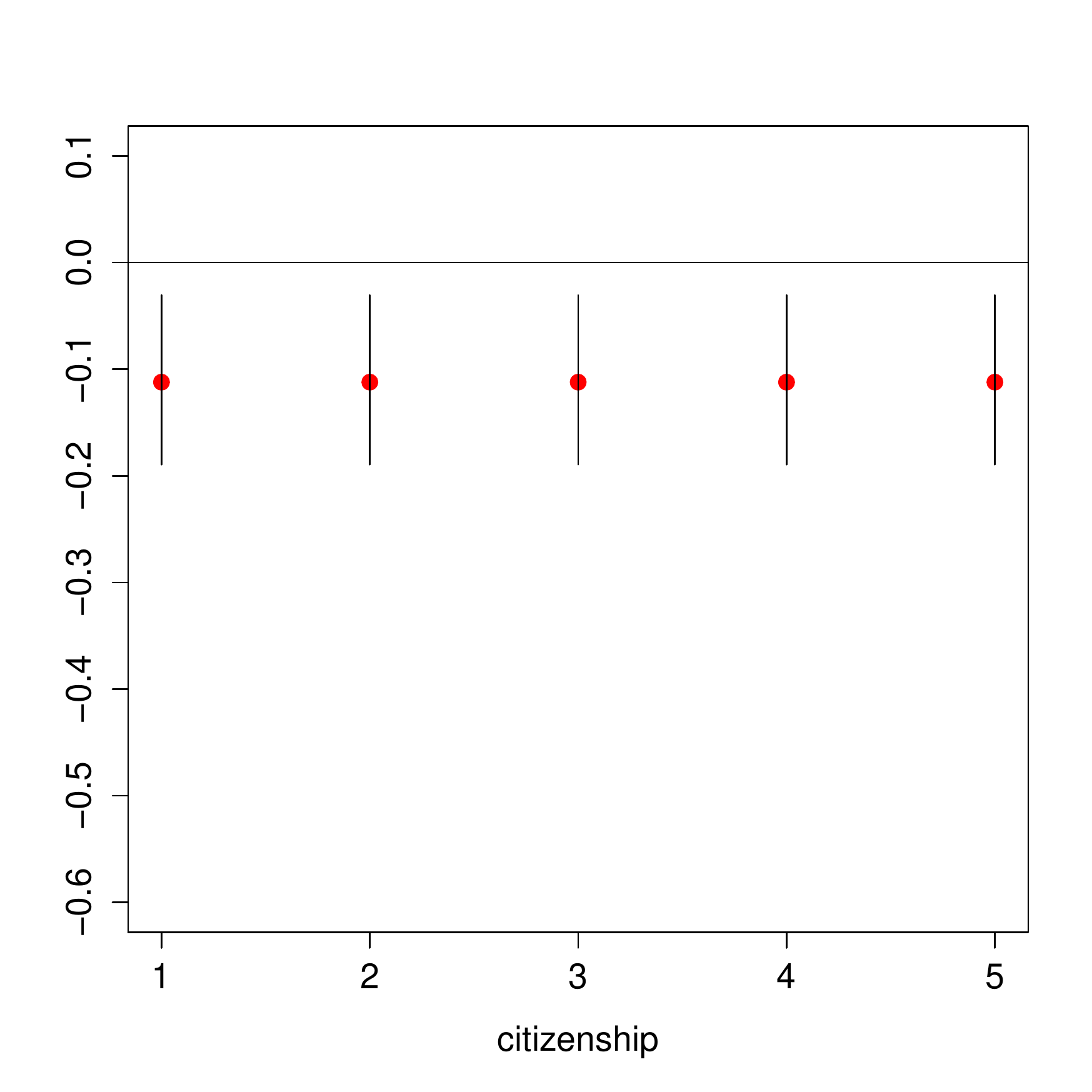} 
	\includegraphics[width=0.32\textwidth]{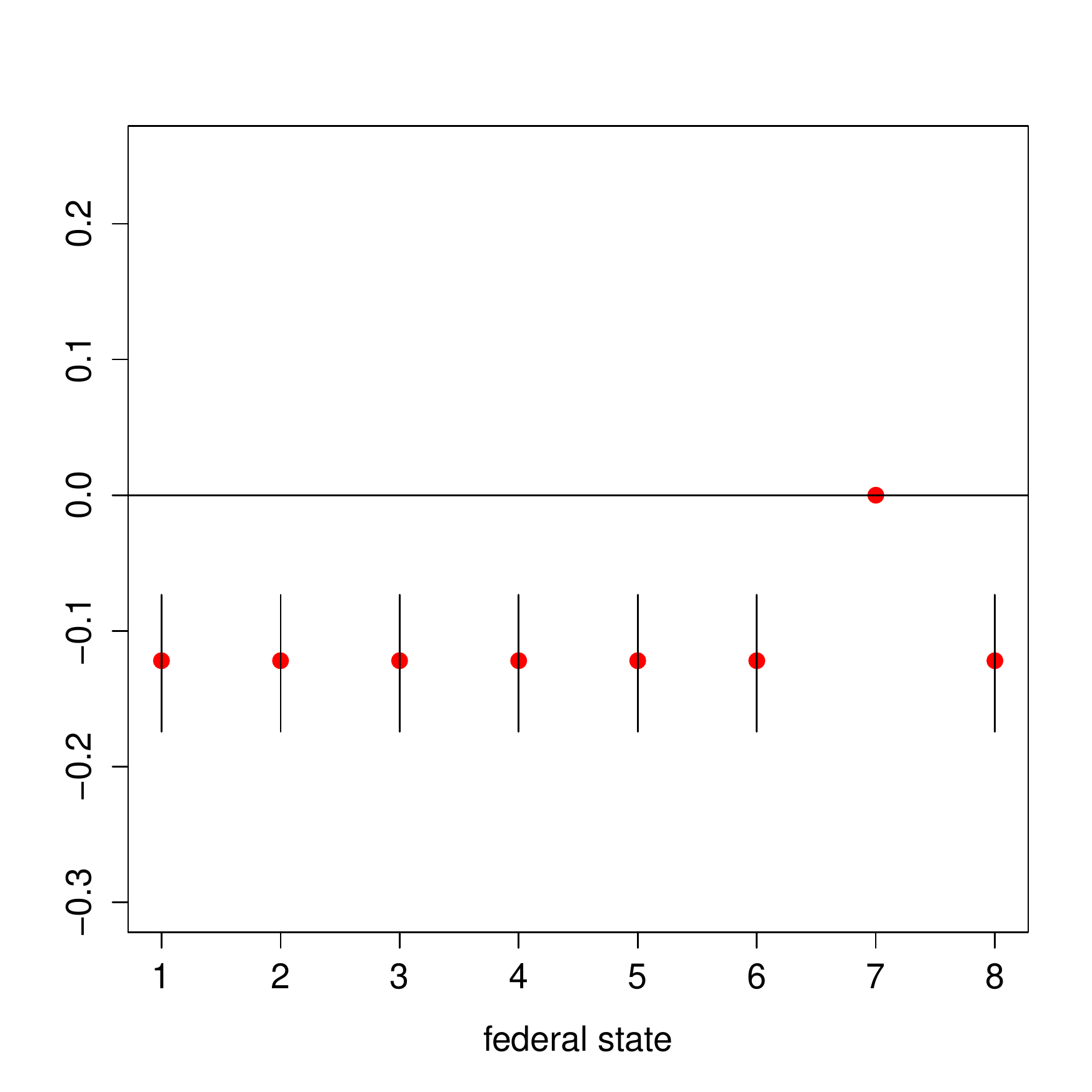} 
	\includegraphics[width=0.32\textwidth]{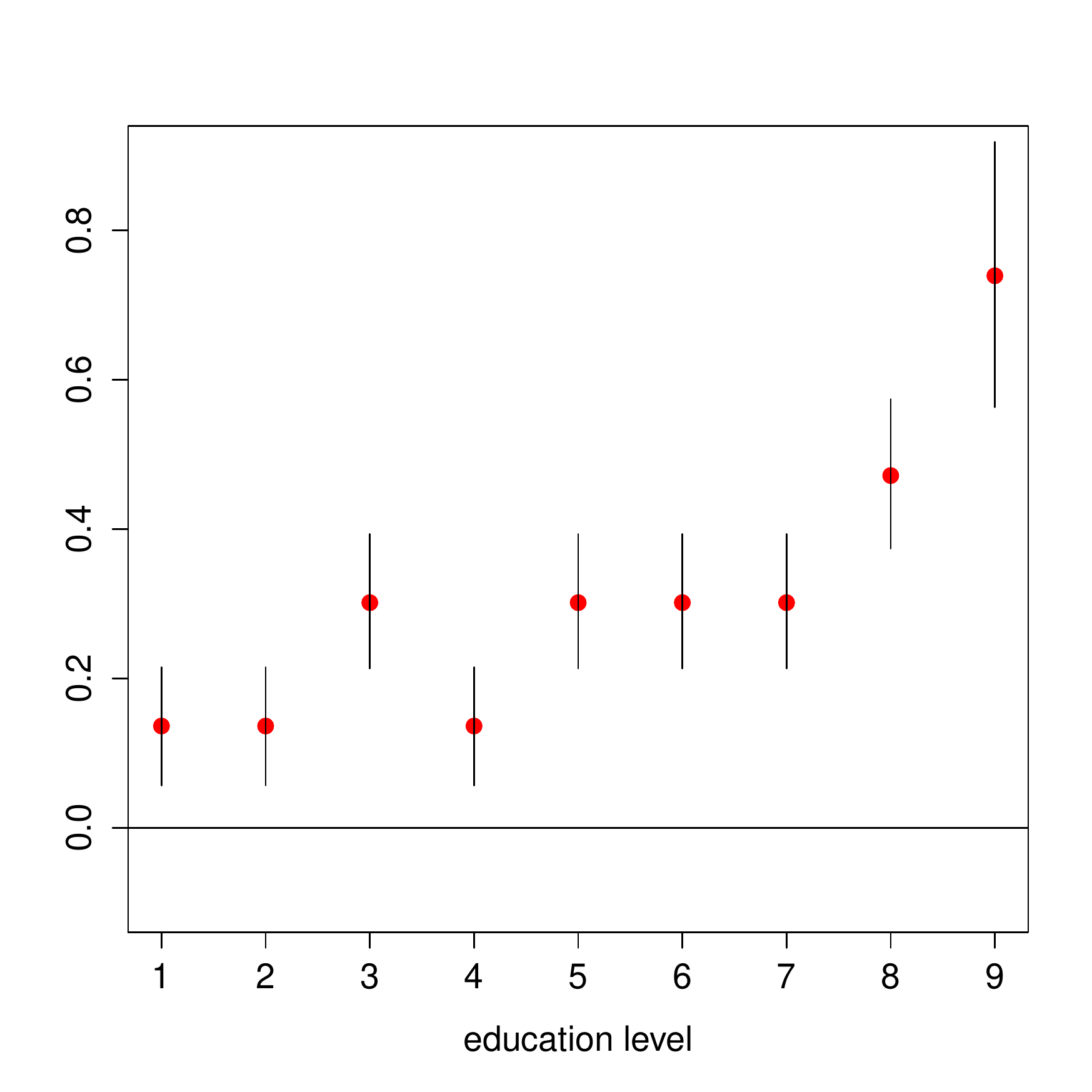} \\
	\includegraphics[width=0.32\textwidth]{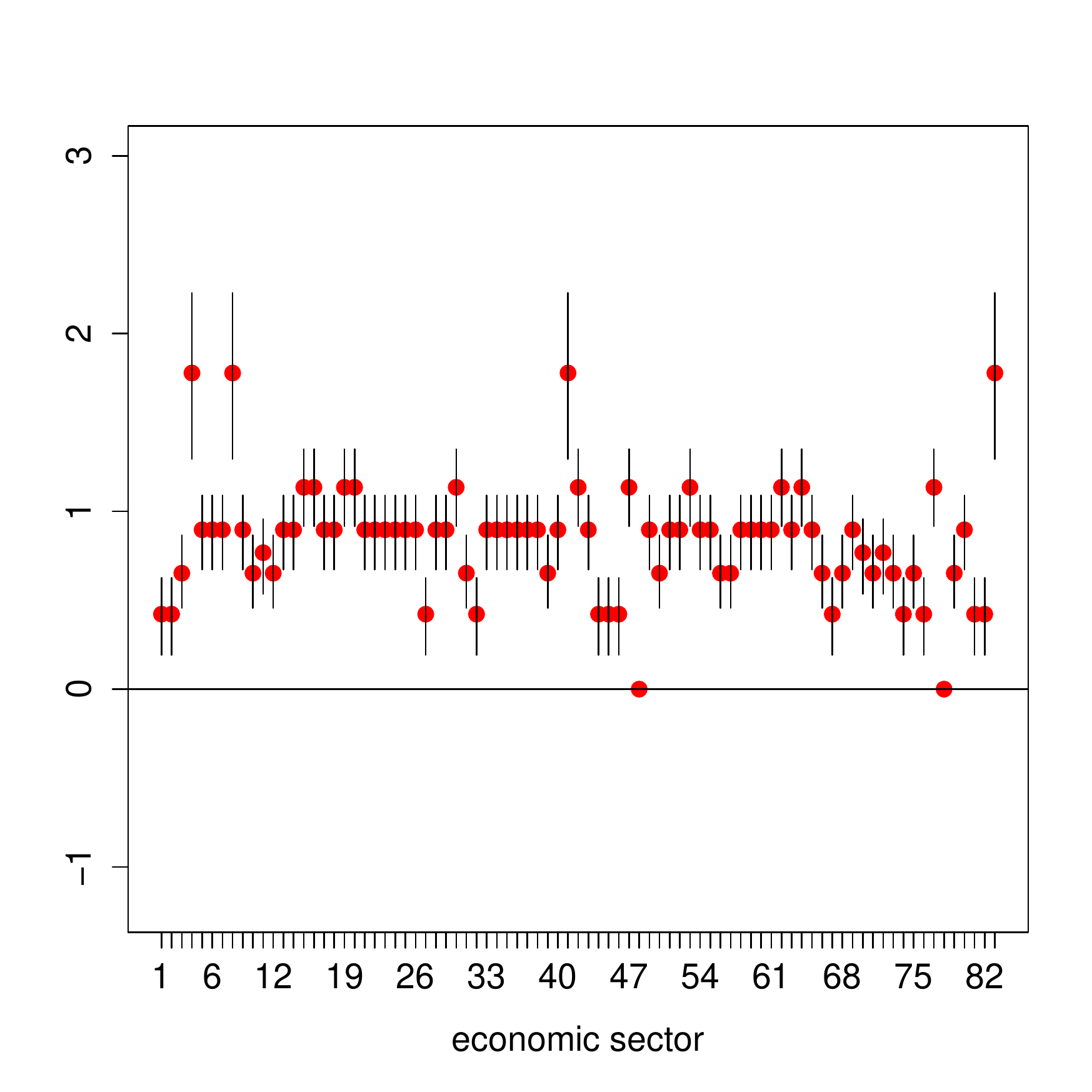} 
	\includegraphics[width=0.32\textwidth]{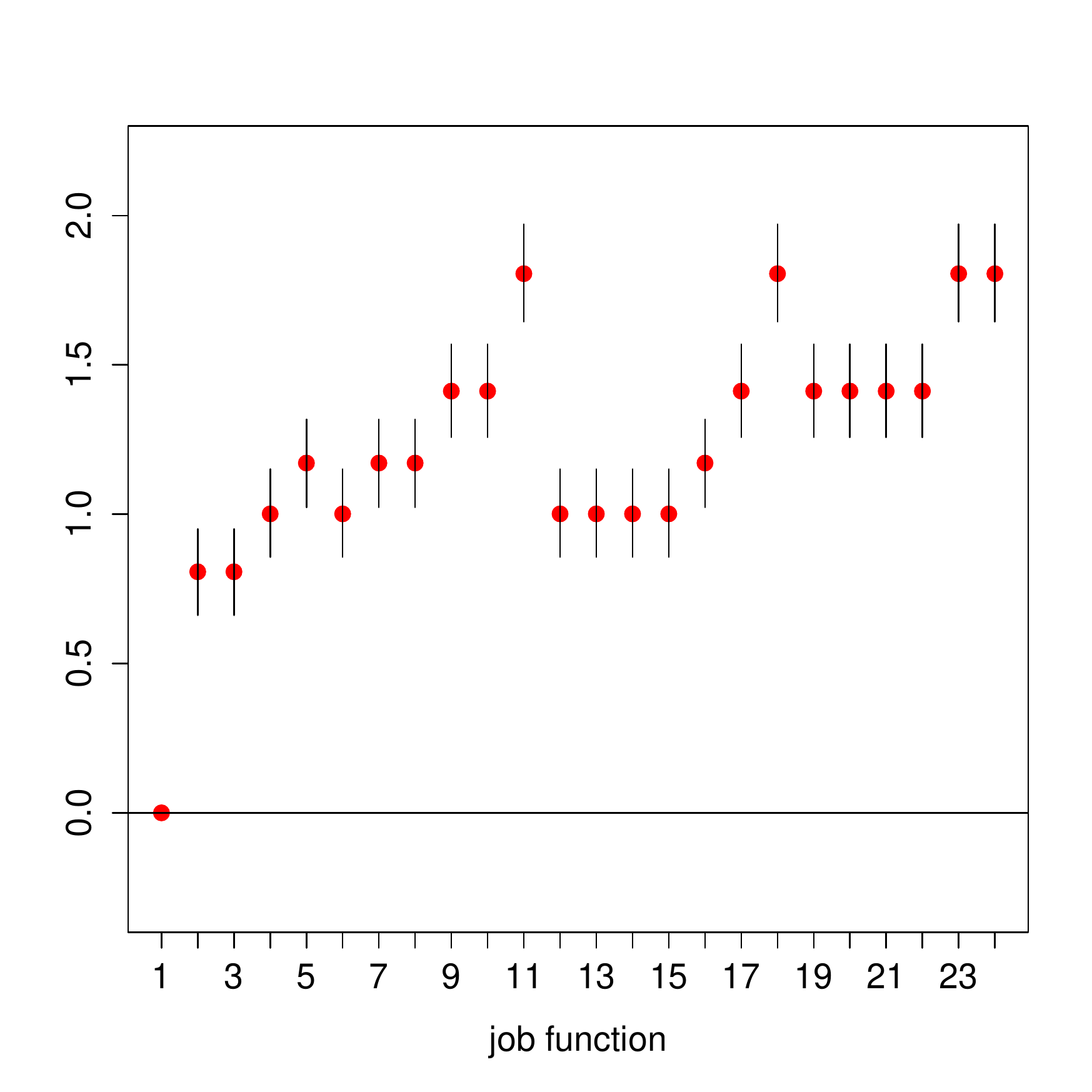} 
	\caption{SILC data, posterior means and  95\% HPD intervals under mixture  prior with $\nu=10^4$, with fixed variance specification.
	}	
	\label{fig:silc_model_c10000_psiFix}
\end{figure}

 For fixed $\psi_j$,  as expected, the number of  effect clusters increases if the component variances decreases. Again, both model selection strategies yield similar clustering results. BICmcmc is smallest for $\nu=10^4$ with 2 effect groups for {\tt citizen}  and {\tt federal state}, 5  for {\tt education}, and 7 and 6 effect groups for {\tt sector} and {\tt job function}, respectively.
The posterior means and the 95 \%HPD intervals of the refitted model are plotted in Figure~\ref{fig:silc_model_c10000_psiFix}.  

\begin{figure}[!ht]\centering
	\includegraphics[width=0.32\textwidth]{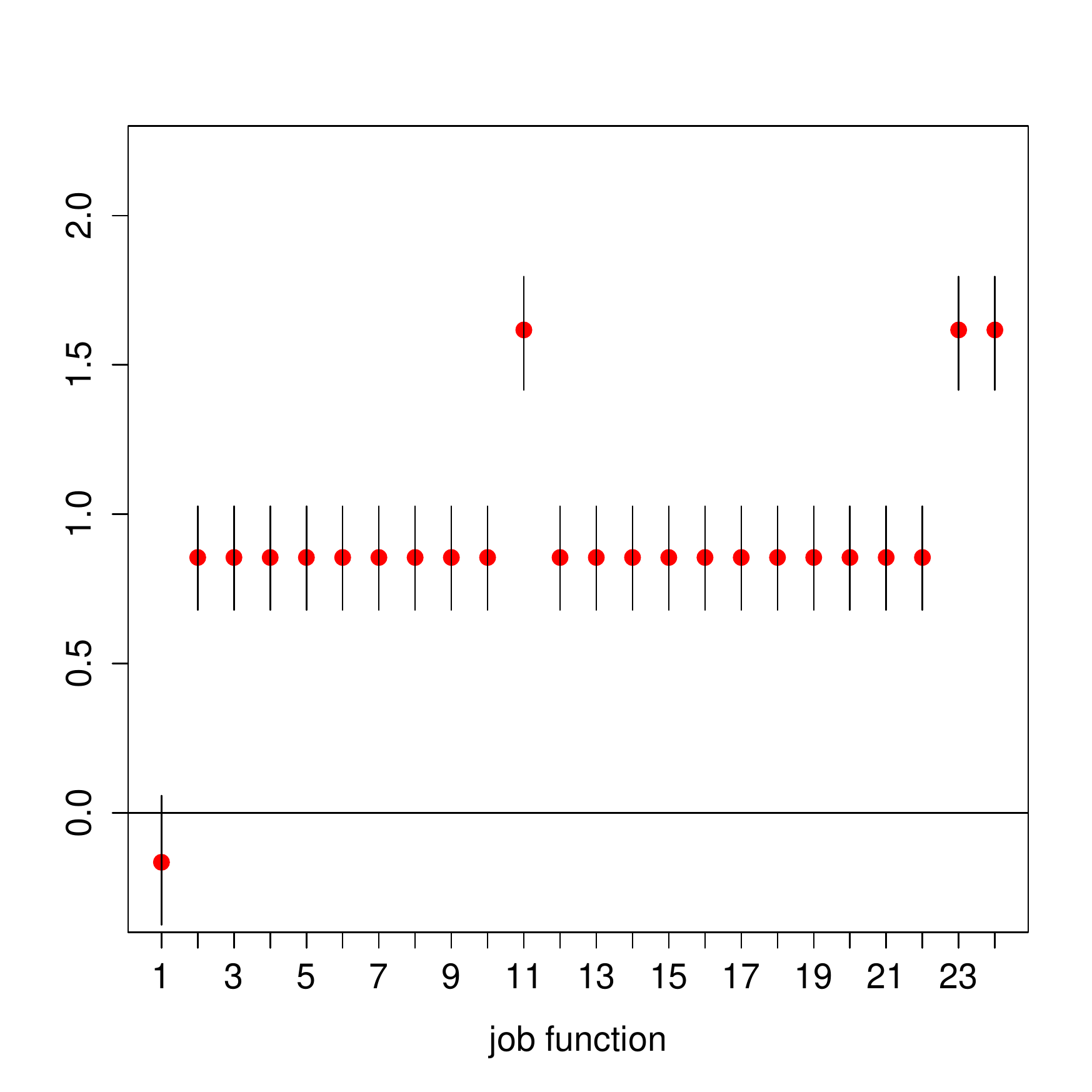} 
	\includegraphics[width=0.32\textwidth]{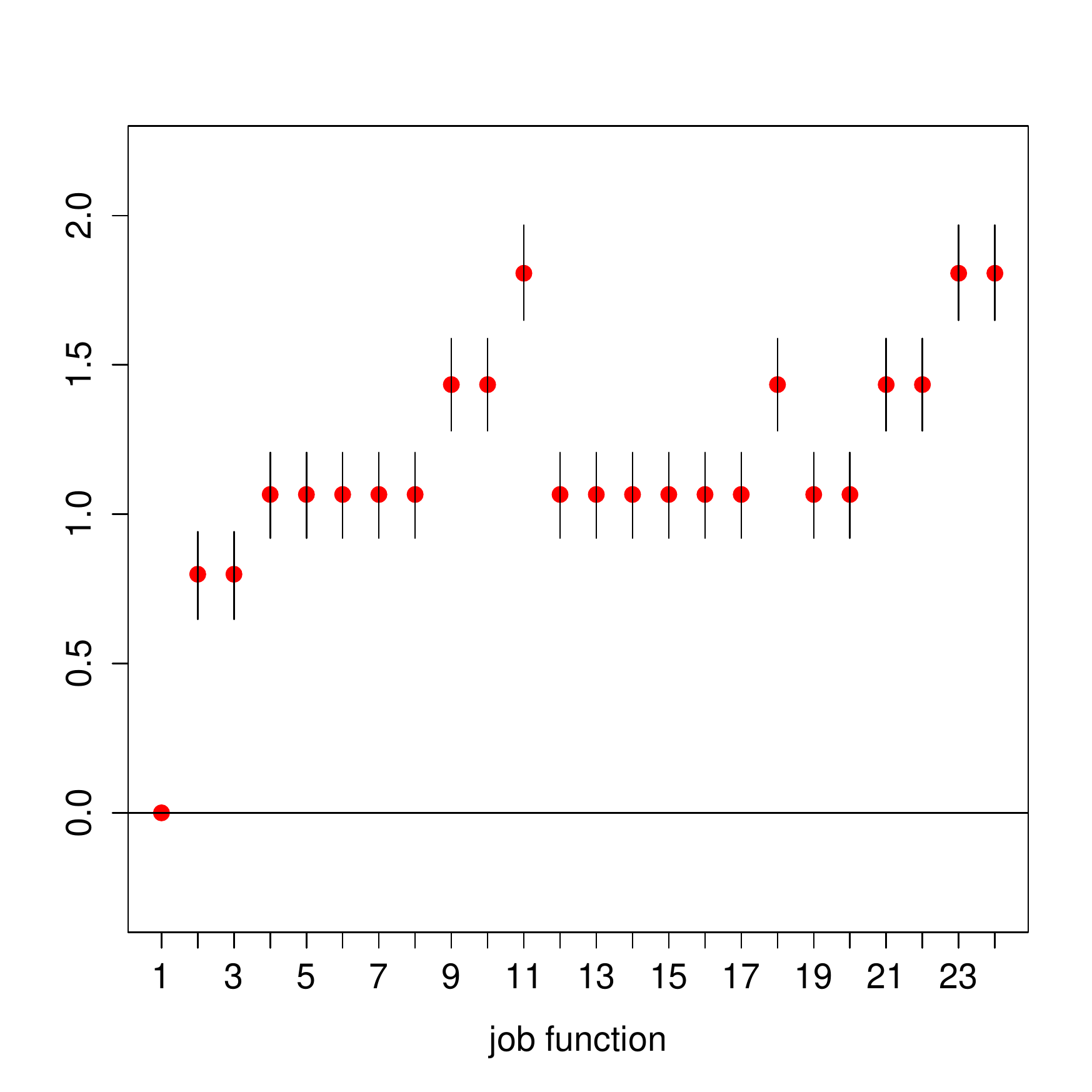} 
	\includegraphics[width=0.32\textwidth]{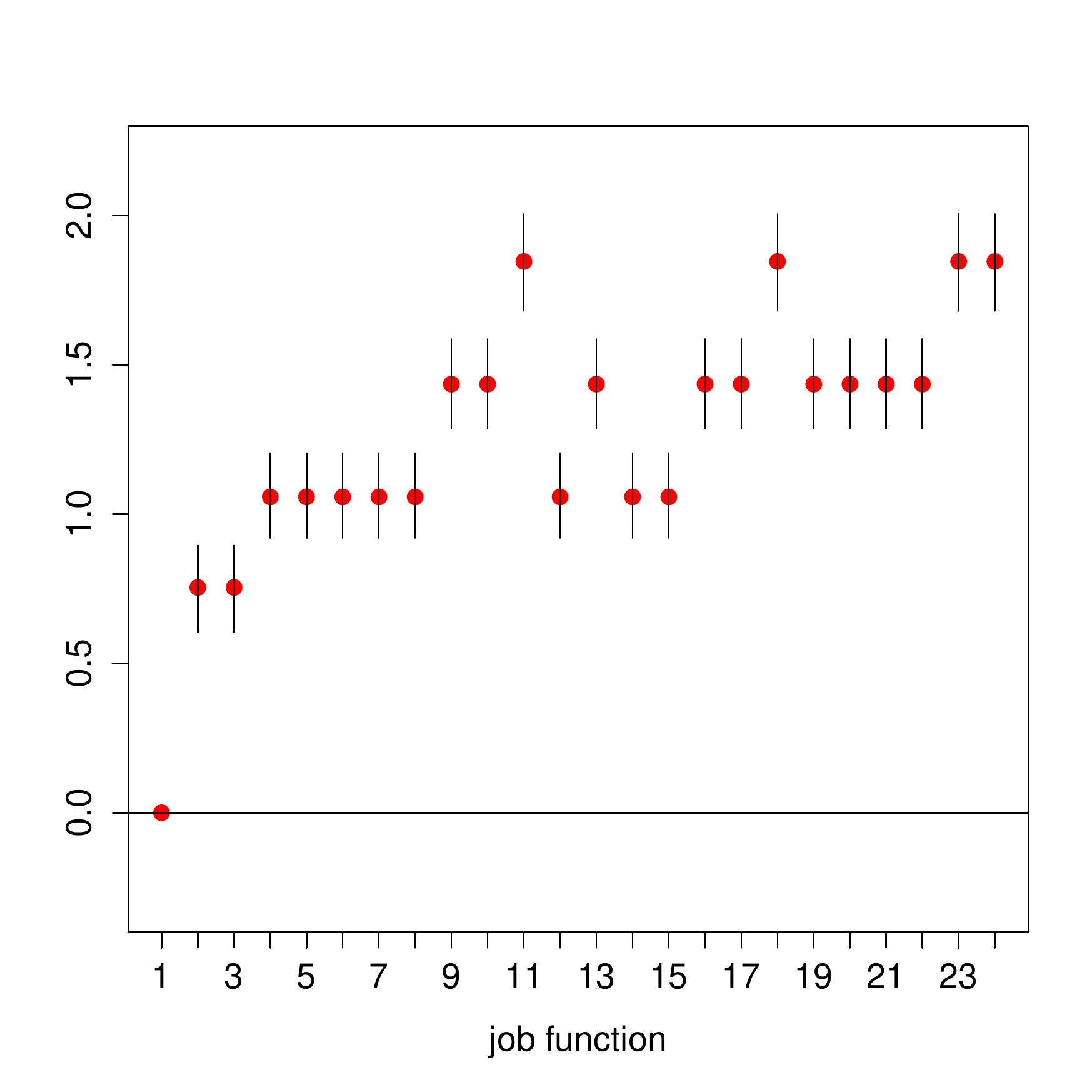}	\\	
	\includegraphics[width=0.32\textwidth]{function_c_10000_fixPsi} 
	\includegraphics[width=0.32\textwidth]{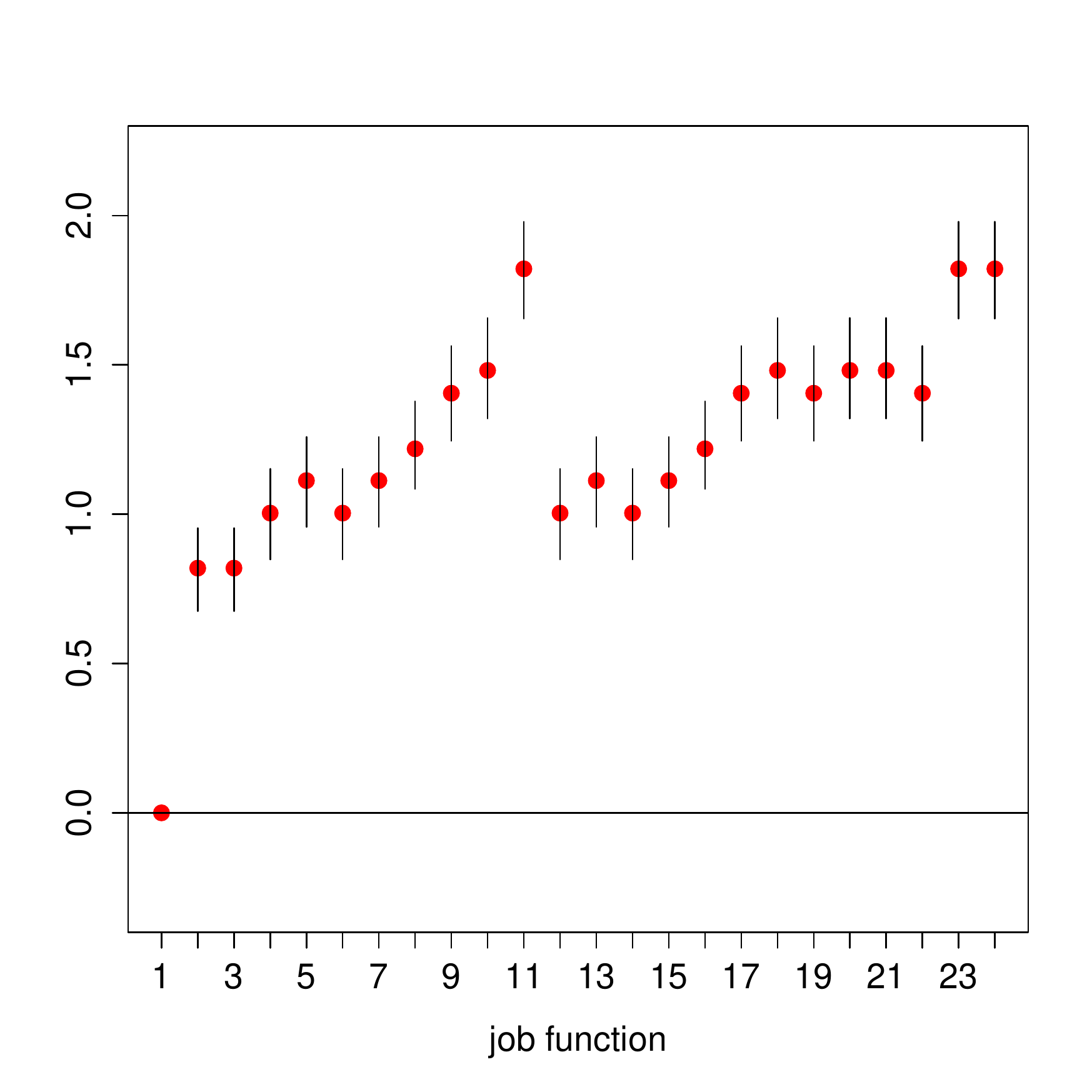} 
	\includegraphics[width=0.32\textwidth]{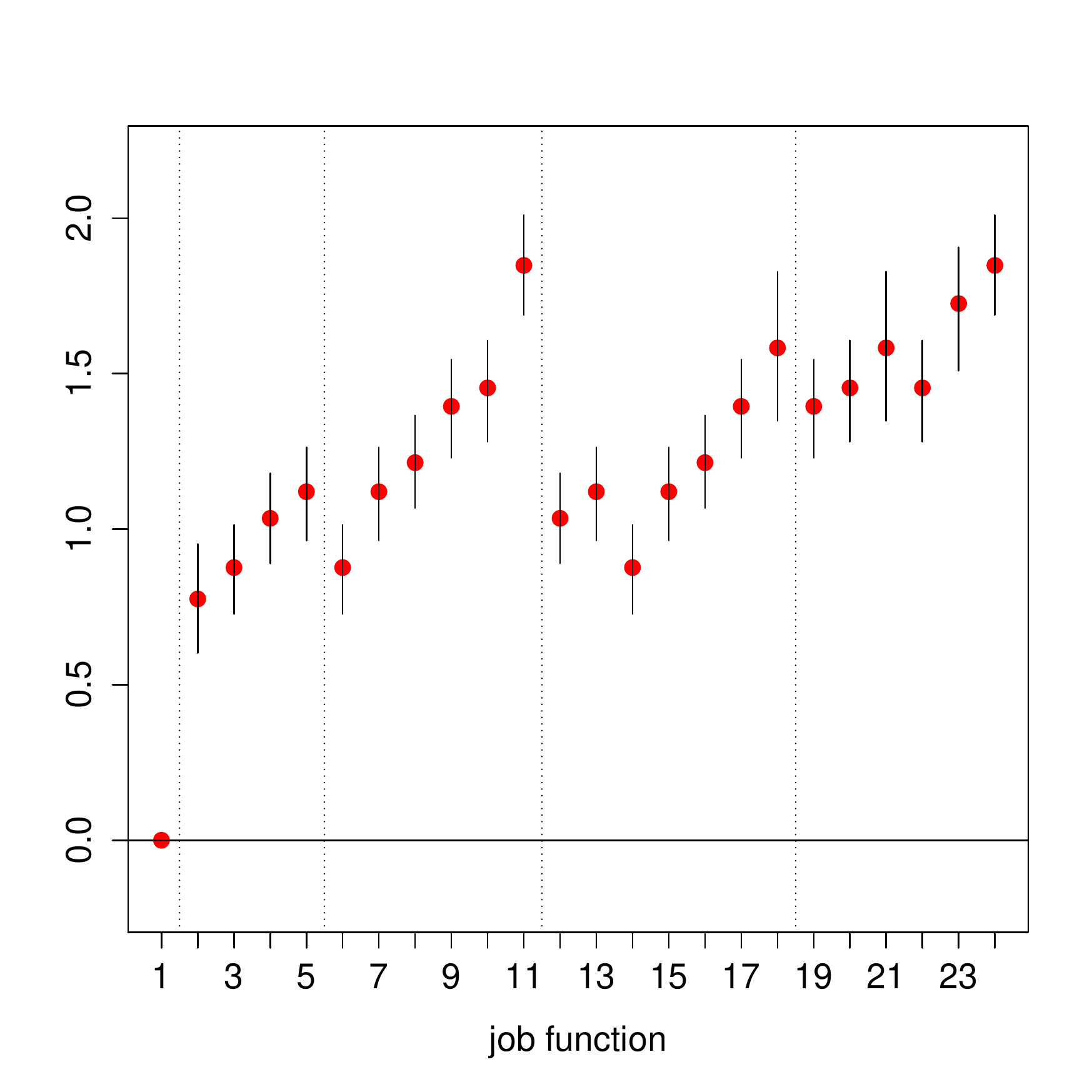} 	
	\caption{SILC data, variable \textit{job function}:  Level estimates with  95\% HPD intervals for various $\nu=10,\ldots,10^6$ (from top-left), by  selecting the   the most frequent model and with fixed component variance $\psi_j$. In the last plot on the bottom-right  the dotted lines indicate the 5 first-level groups, see Table \ref{tab:jobFunction} in the Appendix.}	\label{fig:silc_function}
\end{figure}

To visualize  the cluster solutions for different values of $\nu$, the  estimated effects of   the (refitted) selected models for variable {\tt job function} are plotted in Figure \ref{fig:silc_function}. Obviously  with decreasing  spike variance  the clustering of the level effects gets 'finer'. With a higher resolution  of the effects (e.g. $\nu \geq 10^5$)   an interesting structure is revealed:  as  levels are ordered  by hierarchy function within each contract type   (see Table \ref{tab:jobFunction} in Appendix \ref{app:silc_res})  obviously effects are fused across contract types. This structure would have been missed by using the coarser classificaton level, whereas  on the other hand   even for the very fine resolution with  $\nu =10^6$ the number of estimated effects is less than half compared to the full model.


With a   hyperprior on the component variance $\psi_j$ the selected models are very sparse and  the number of effect clusters is almost constant, see Table \ref{tab:silc}. This is in  agreement with the results from the  simulation  study and  the considerably higher values of BICmcmc indicate that a prior with fixed component variances should be prefered.

\clearpage
\section{Discussion} \label{sec:discussion}

In this paper, we proposed  to specify finite normal mixture priors on the level effects of a categorical predictor to obtain a sparse representation of  these effects. 
The mixture specification allows to shrink non-zero effects to different non-zero locations and introduces a natural clustering of the level effects. Level effects assigned to the same mixture component are  fused, i.e.~their effects are  replaced by the same  joint effect. The number of components as well as their locations are treated as unknown and estimated from the data. A sparse prior on the mixture weights helps to avoid unnecessary splitting of non-empty components and to concentrate the posterior distribution on a sparse cluster solution. 
The number of estimated level groups  can be guided by the size of  variance of the  mixture components, with a smaller variances inducing a larger number of estimated effect groups.

We noted that surprisingly the specification of a hyperprior on the component covariances  did not work well. In contrast to the common clustering of data, we  aim at clustering of  regression effects, which are not fixed but have to be estimated from the data. Assigning an effect to a mixture component corresponds to selecting a particular prior distribution for its estimation and hence has  an impact on its  value in the next model based clustering step. Thus additional  uncertainty  is introduced which  results in the estimation of large component variances and only  few effect groups.  
Therefore, we recommend to fix the variance of mixture component and  investigate the resolution of level effects with different values. To select the final model model choice criteria can be used.
A strength of our approach is that the spike variance specification can vary across the variables, which  allows the researcher to obtain a 'finer' clustering for effects of particular interest.



We investigated two different model selection strategies, either  to select  model sampled  most frequently (`most')  or  to apply   PAM to the matrix of posterior  inclusion probabilities and select the final model using the silhoutte coefficient (`pam'). 
Both strategies 
 have shown to perform  similar. An advantage of the 'most' strategy  is, that also a one-group solution can be selected, which is not possibel for the  `pam'  strategy, but  the later is robust against the switching of single effects between groups.


The approach works well even if the number of categories is high, e.g.~around 100.  For Gaussian response regression models the computational effort is low as  a standard Gibbs sampling 
scheme can be used  for MCMC estimation. However the method is not at all restricted to Gaussian regression models. It can be easily implemented  as an  ''add-on''  to an MCMC sampling scheme 
for any regression type model  with  a multivariate Normal prior on the regression effects, as  in each MCMC iteration only  the steps  for  model based clustering as well as the  update of the prior parameters of the regression effects is required.

\section*{Acknowledgements}
This research was financially supported by the Austrian Science Fund FWF (P25850, V170 and P28740) and the Austrian National Bank (Jubil\"aumsfond  14663). We want to thank Bettina Gr\"un from the Department of Applied Statistics, Johannes Kepler University Linz, for many  fruitful discussions.

\appendix
\section{Appendix}

\subsection{MCMC sampling} \label{app:MCMC}
Let $\boldsymbol{b}_0(\boldsymbol{S})$ and $\boldsymbol{B}_0(\boldsymbol{S})$ denote  the mean vector and the covariance matrix of the vector of all regression effects $\beta_{jk}$ conditional on their component indicators $S_{jk}$, i.e.
$$ \bbeta|\boldsymbol{S} \sim \mathcal{N}(\boldsymbol{b}_0(\boldsymbol{S}),\boldsymbol{B}_0(\boldsymbol{S})),$$
where $\boldsymbol{b}_0(\boldsymbol{S}) = (0,\mu_{1S_{11}},\ldots,\mu_{1 S_{1L_1}},\ldots,\mu_{JS_{J1}},\ldots,\mu_{JS_{JL_J}})$ and $ \boldsymbol{B}_0(\boldsymbol{S})$ is a diagonal matrix with entries $(\psi_0,\psi_{1S_{11}},\ldots,\psi_{1 S_{1L_1}},\ldots,\psi_{JS_{J1}},\ldots,\psi_{JS_{JL_J}})$. Posterior inference using MCMC sampling iterates the following steps:

\begin{itemize}
	\item[] {\bf Regression steps}
\begin{enumerate}
	\item Sample the regression coefficients $\bbeta$ conditional on  $\boldsymbol{S}$ from the normal posterior $\mathcal{N}(\boldsymbol{b}_N, \boldsymbol{B}_N)$, where
	\begin{align*}
			\boldsymbol{B}_N & = \sigma^2 (\mathbf{X}'\mathbf{X} + \sigma^2 \boldsymbol{B}_0(\boldsymbol{S})^{-1})^{-1} \\
			\boldsymbol{b}_N & = \boldsymbol{B}_N (\mathbf{X}'y / \sigma^2 + \boldsymbol{B}_0(\boldsymbol{S})^{-1} \boldsymbol{b}_0(\boldsymbol{S})).  
	\end{align*}
	\item Sample the error variance $\sigma^2$ from its full conditional posterior distribution $\mathcal{G}^{-1}(s_N, S_N)$, where 
	\begin{align*}
			s_N & = s_0 + N/2 \\
			S_N & = S_0 + \frac{1}{2}(y-\Xv \bbeta)'(y-\Xv\bbeta).
	\end{align*}
\end{enumerate}	
	\item[] {\bf Model based clustering steps}
	\begin{enumerate}
	\setcounter{enumi}{3}
	\item For $j=1,...,J$ sample the component weights $\boldeta_j$ from the Dirichlet distribution, $Dir(e_{j0},e_{j1},\ldots,e_{jL_j})$, where 
	$$e_{jl} =e_0+N_{jl}, \quad l=0,\ldots,L$$ 
	and $N_{jl}$ is the number of regression coefficients $\beta_{jk}$ of covariate $j$ assigned to   mixture component $l$.	
	\item For $j=1,...,J$; $l=1,\dots L_j$ sample the mixture component means $\mu_{jl}$ from their normal posterior $\mathcal{N}(m_{jl},M_{jl})$, where 
	\begin{align*}
			M_{jl} & = (N_{jl}/\psi_{j} + M_{0j}^{-1})^{-1},\\
			m_{jl} & = M_{jl}(N_{jl} \bar{\beta}_{jl}/\psi_{j} +M_{0j}^{-1} m_{0j})
	\end{align*}
	and $\bar{\beta}_{jl}$ is the mean of all  elements of $\bbeta_j$ assigned to component $l$.
	\item If a hyperprior is specified on  the mixture component variances $\psi_{j}$, sample  $\psi_{j}$ for $j=1,\dots, J$
		 from its inverse Gamma posterior $\mathcal{G}^{-1}(g_{jN},G_{jN})$, where
	\begin{align*}
			g_{jN} & = g_0 + c_j/2 \\
			G_{jN} & = G_0 + \frac{1}{2} \sum_{k:S_{jk}=l} \sum_{l=0}^{L_j} (\beta_{jk}-\mu_{jl})^2.
	\end{align*}
	\item Sample the vector of the latent allocation indicators $\boldsymbol{S}$ from the full conditional posterior 
	$$ P(S_{jh}=l|\beta_{jh},\boldsymbol{\mu}_j,\boldsymbol{\psi}_j) \propto \eta_{jl} f_{\mathcal{N}}(\beta_{jh}|\mu_{jl},\psi_{j}) \qquad j=1,\dots, J; h
	=1,\dots, L_j$$
	and update $\boldsymbol{b}_0(\boldsymbol{S}), \boldsymbol{B}_0(\boldsymbol{S}), N_{jl}$ and $ \bar{\beta}_{jl} $ for $l=1,\dots L_j$.
\end{enumerate}
\end{itemize}

\subsection{Definitions}

\subsubsection{Silhouette coefficient} \label{app:silhouette}

The silhouette coefficient in \cite{rou:sil} is defined as follows. Let $i$ be any object in the data set and $A$ is the cluster to which it has been assigned. If cluster $A$ contains other objects apart from $i$, then $a(i)$ is the average dissimilarity of $i$ to all other objects of $A$. $d(i,C)$ is the average dissimilarity of $i$ to all objects in cluster $C$ which represents any cluster different from $A$. Compute $d(i,C)$ for all clusters $C \ne A$ and denote by $b(i) = \min \limits_{C \ne A} d(i,C)$. The silhouette coefficients is then computed as
\begin{align*}
	s(i) = \frac{b(i)-a(i)}{\max(a(i),b(i)).}
\end{align*}

\subsubsection{Adjusted Rand index} \label{app:randindex}

The adjusted Rand index \citep{hub-ara:com} is a  form of the Rand index \citep{ran:obj} which is  adjusted for chance agreement. If  $n$ is the number of elements and $\Xv = \{X_1,X_2,...X_r\}$ and $\boldsymbol{Y} = \{Y_1,Y_2,...,Y_s \}$ are two clusterings of these elements, the adjusted Rand index is defined as
\begin{align*}
	\text{AR} = \frac{\sum_{ij} \binom{n_{ij}}{2} - [\sum_i \binom{a_i}{2} \sum_j \binom{b_j}{2}] / \binom{n}{2} }
	{\frac{1}{2} [\sum_i \binom{a_i}{2} + \sum_j \binom{b_j}{2}] - [\sum_i \binom{a_i}{2} \sum_j \binom{b_j}{2}] / \binom{n}{2}},
\end{align*}
where $a_i$  and $b_j$ are  the number of objects in $X_i$  and $Y_j$, respectively and $n_{ij}$ is the number of objects in  $X_i \cap Y_j$.

\subsection{Further simulation results} \label{app:sim_res}
We report simulation results for covariates 1 to 3 of the simulation study in Tables \ref{tab:ev_res_var1} to \ref{tab:ev_res_var3}.

\begin{table}[ht]
	\centering
	\begin{tabular}{|c|c|c|cc|cc|cc|cc|cc|}
		\hline
		&$\nu$ & freq&  \multicolumn{2}{c|}{groups}  & \multicolumn{2}{c|}{AR} & \multicolumn{2}{c|}{Error} & \multicolumn{2}{c|}{FPR} & \multicolumn{2}{c|}{FNR} \\
		&  &  & most & pam  & most & pam & most & pam & most & pam & most & pam\\ 
		\hline
		fixed & 10 & 8324 & 2.8 & 3.0 & 0.89 & 0.99 & 0.07 & 0.00 & 0.00 & 0.01 & 0.07 & 0.00 \\ 
		& $10^2$ & 14466 & 3.0 & 3.0 & 1.00 & 1.00 & 0.00 & 0.00 & 0.00 & 0.00 & 0.00 & 0.00 \\ 
		& $10^3$ & 14844 & 3.0 & 3.0 & 1.00 & 1.00 & 0.00 & 0.00 & 0.00 & 0.00 & 0.00 & 0.00 \\ 
		& $10^4$ & 14847 & 3.0 & 3.0 & 1.00 & 1.00 & 0.00 & 0.00 & 0.00 & 0.00 & 0.00 & 0.00 \\ 
		& $10^5$ & 14604 & 3.2 & 3.2 & 0.97 & 0.97 & 0.02 & 0.02 & 0.04 & 0.05 & 0.00 & 0.00 \\ 
		& $10^6$ & 13673 & 4.3 & 4.4 & 0.78 & 0.77 & 0.15 & 0.15 & 0.28 & 0.30 & 0.00 & 0.00 \\ 
		\hline 
		\hline
		random & 10 & 8054 & 2.8 & 3.0 & 0.90 & 0.99 & 0.07 & 0.00 & 0.00 & 0.00 & 0.06 & 0.00 \\ 
		& $10^2$ & 13940 & 3.0 & 3.0 & 1.00 & 1.00 & 0.00 & 0.00 & 0.00 & 0.00 & 0.00 & 0.00 \\ 
		& $10^3$ & 14250 & 3.0 & 3.0 & 1.00 & 1.00 & 0.00 & 0.00 & 0.00 & 0.00 & 0.00 & 0.00 \\ 
		& $10^4$ & 14343 & 3.0 & 3.0 & 1.00 & 1.00 & 0.00 & 0.00 & 0.00 & 0.00 & 0.00 & 0.00 \\ 
		& $10^5$ & 14308 & 3.0 & 3.0 & 1.00 & 1.00 & 0.00 & 0.00 & 0.00 & 0.00 & 0.00 & 0.00 \\ 
		& $10^6$ & 14308 & 3.0 & 3.0 & 1.00 & 1.00 & 0.00 & 0.00 & 0.00 & 0.00 & 0.00 & 0.00 \\ 
		\hline
	\end{tabular}
	\caption{Model selection results: \textbf{Var1}, $10$ categories, true number of groups is $3$.} 
	\label{tab:ev_res_var1}
\end{table}

\begin{table}[ht]
	\centering
	\begin{tabular}{|c|c|c|cc|cc|cc|cc|cc|}
		\hline
		&$\nu$ & freq&  \multicolumn{2}{c|}{groups}  & \multicolumn{2}{c|}{AR} & \multicolumn{2}{c|}{Error} & \multicolumn{2}{c|}{FPR} & \multicolumn{2}{c|}{FNR} \\  
		&  &  & most & pam  & most & pam & most & pam & most & pam & most & pam\\ 
		\hline
		fixed & 10 & 14047 & 2.0 & 2.0 & 1.00 & 1.00 & 0.00 & 0.00 & 0.00 & 0.00 & 0.00 & 0.00 \\ 
		& $10^2$ & 14601 & 2.0 & 2.0 & 1.00 & 0.99 & 0.00 & 0.00 & 0.00 & 0.00 & 0.00 & 0.00 \\ 
		& $10^3$ & 14970 & 2.0 & 2.0 & 1.00 & 0.99 & 0.00 & 0.00 & 0.00 & 0.00 & 0.00 & 0.00 \\ 
		& $10^4$ & 14929 & 2.0 & 2.0 & 0.99 & 0.99 & 0.00 & 0.00 & 0.00 & 0.01 & 0.00 & 0.00 \\ 
		& $10^5$ & 14263 & 2.4 & 2.5 & 0.75 & 0.70 & 0.10 & 0.13 & 0.17 & 0.21 & 0.00 & 0.00 \\ 
		& $10^6$ & 14095 & 3.4 & 3.5 & 0.30 & 0.28 & 0.33 & 0.35 & 0.52 & 0.54 & 0.00 & 0.00 \\
		\hline
		random & 10 & 13789 & 2.0 & 2.0 & 0.99 & 0.99 & 0.00 & 0.00 & 0.00 & 0.00 & 0.00 & 0.00 \\ 
		& $10^2$ & 13651 & 2.0 & 2.0 & 1.00 & 1.00 & 0.00 & 0.00 & 0.00 & 0.00 & 0.00 & 0.00 \\ 
		& $10^3$ & 13711 & 2.0 & 2.0 & 1.00 & 1.00 & 0.00 & 0.00 & 0.00 & 0.00 & 0.00 & 0.00 \\ 
		& $10^4$ & 13856 & 2.0 & 2.0 & 1.00 & 1.00 & 0.00 & 0.00 & 0.00 & 0.00 & 0.00 & 0.00 \\ 
		& $10^5$ & 13915 & 2.0 & 2.0 & 1.00 & 1.00 & 0.00 & 0.00 & 0.00 & 0.00 & 0.00 & 0.00 \\ 
		& $10^6$ & 13678 & 2.0 & 2.0 & 1.00 & 1.00 & 0.00 & 0.00 & 0.00 & 0.00 & 0.00 & 0.00 \\
		\hline 
	\end{tabular}
	\caption{Model selection results: \textbf{Var2}, $10$ categories, true number of groups is $2$.} 
	\label{tab:ev_res_var2}
\end{table}

\begin{table}[ht]
	\centering
	\begin{tabular}{|c|c|c|cc|cc|cc|cc|cc|}
		\hline
		&$\nu$ & freq&  \multicolumn{2}{c|}{groups}  & \multicolumn{2}{c|}{AR} & \multicolumn{2}{c|}{Error} & \multicolumn{2}{c|}{FPR} & \multicolumn{2}{c|}{FNR} \\  
		&  &  & most & pam  & most & pam & most & pam & most & pam & most & pam\\ 
		\hline
		fixed & 10 & 9154 & 1.1 & 2.0 & 0.86 & 0.00 & 0.01 & 0.10 & 0.03 & 0.20 & - & - \\ 
		& $10^2$ & 12591 & 1.1 & 2.1 & 0.90 & 0.00 & 0.01 & 0.14 & 0.02 & 0.26 & - & - \\ 
		& $10^3$ & 11897 & 1.8 & 2.0 & 0.26 & 0.00 & 0.23 & 0.28 & 0.33 & 0.41 & - & - \\ 
		& $10^4$ & 12003 & 3.1 & 3.3 & 0.00 & 0.00 & 0.47 & 0.50 & 0.63 & 0.67 & - & - \\ 
		& $10^5$ & 12402 & 4.9 & 4.6 & 0.00 & 0.00 & 0.63 & 0.64 & 0.82 & 0.81 & - & - \\ 
		& $10^6$ & 12709 & 7.1 & 5.5 & 0.00 & 0.00 & 0.74 & 0.68 & 0.92 & 0.85 & - & - \\ 
		\hline
		random & 10 & 9027 & 1.1 & 2.0 & 0.87 & 0.00 & 0.01 & 0.10 & 0.03 & 0.21 & - & - \\ 
		& $10^2$ & 10091 & 2.0 & 2.0 & 0.02 & 0.00 & 0.10 & 0.10 & 0.20 & 0.20 & - & - \\ 
		& $10^3$ & 10079 & 2.0 & 2.0 & 0.01 & 0.00 & 0.10 & 0.10 & 0.20 & 0.20 & - & - \\ 
		& $10^4$ & 10043 & 2.0 & 2.0 & 0.02 & 0.00 & 0.10 & 0.10 & 0.20 & 0.20 & - & - \\ 
		& $10^5$ & 10132 & 2.0 & 2.0 & 0.01 & 0.00 & 0.10 & 0.10 & 0.20 & 0.20 & - & - \\ 
		& $10^6$ & 10162 & 2.0 & 2.0 & 0.00 & 0.00 & 0.10 & 0.10 & 0.20 & 0.20 & - & - \\ 
		\hline
	\end{tabular}
	\caption{Model selection results: \textbf{Var3}, $10$ categories, true number of groups is $1$, i.e.~all effects should be fused to the baseline.} 
	\label{tab:ev_res_var3}
\end{table}

\subsection{SILC data} \label{app:silc_res}

Table \ref{tab:jobFunction} describes the two-level classification scheme of the variable {\tt job function} and the frequencies of the categories.
	\begin{footnotesize}
	\begin{table}[h!]	
	\centering	
		\begin{tabular}{|c|c|c|}	\hline 
			Level I (contract type)	& Level II (skills) & Number of observations\\ 	
			\hline 
			apprentice	& for white-collar worker& 114\\ 
						&  for   blue-collar worker&66 \\ \hline 
			blue-collar worker	&  unskilled worker& 143\\ 
								& semi-skilled worker& 413\\
								& skilled worker& 555\\	
								& foreman& 83\\							
			\hline
			white-collar worker &  simple activities& 85\\ 
								& trained abilities/tasks& 300\\
								& medium abilities/tasks& 543\\
								& superior activities/tasks& 388\\
								& highly qualified activities& 250\\
								& leading activities& 358\\
			\hline
			contract staff &  simple activities& 6\\ 
							&  craftsmanship activities & 13\\
							&  auxiliary activities & 8\\
							&  trained abilities/tasks & 31\\
							&medium abilities/tasks &57\\
							&superior activities/tasks &61\\
							& highly qualified or leading activities& 21\\
			\hline
			officials & craftsmanship activities& 10\\ 
						 & auxiliary activities& 3\\
						& trained abilities/tasks & 27\\
						&  medium abilities/tasks & 137\\
						&superior activities/tasks &112\\
						& highly qualified or leading activities& 81\\	
			\hline 
			\textbf{5} & \textbf{25} &3865\\ \hline
		\end{tabular}
		\caption{SILC data Austria 2010, variable {\tt job function}: Five categories on the first level, 25 categories on the second level. 
			} \label{tab:jobFunction}
	\end{table}
\end{footnotesize}



\clearpage
\bibliography{mixture}

\end{document}